\documentclass[11pt]{article}
\usepackage[dvips]{color}
\usepackage{amssymb}
\usepackage{amsmath}
\usepackage{amsfonts}
\usepackage{url}
\usepackage{bm}
\usepackage{longtable}
\usepackage{textgreek}

\usepackage{ifpdf}

\usepackage[pdftex]{graphicx}
\usepackage{authblk}

\usepackage{lineno}

\def\vk{{\bf k}}
\def\vr{{\bf r}}
\def\vR{{\bf R}}
\DeclareMathOperator{\Tr}{Tr}

\begin{document}

		\title{Electronic correlations in dense iron: from moderate pressure to Earth's core conditions}
		\author{Leonid V. Pourovskii}
		
		\affil{\small{CPHT, CNRS, Ecole Polytechnique, IP Paris, Route de Saclay, 91128 Palaiseau, France \\
			Coll\`{e}ge de France, 11 place Marcelin Berthelot, 75005 Paris, France}}
		
		\maketitle
		
	\begin{abstract}
		\small{We discuss the role of dynamical many-electron effects in the physics of iron and iron-rich
		solid alloys under applied pressure on the basis of recent {\it ab initio} studies
		employing the dynamical mean-field theory (DMFT). We review in details two particularly interesting regimes: first, a moderate pressure range  up to 60 GPa and, second, the ultra-high pressure of about 360 GPa expected inside the solid inner core of Earth. 
		
	Electronic correlations in iron under the moderate pressure of several tens GPa are discussed in the first section. 
	DMFT-based methods predict an enhancement of electronic correlations at the pressure-induced body-centered cubic $\alpha$ $\to$ hexagonal close-packed $\epsilon$ phase transition.
		In particular, the electronic effective mass, scattering rate and electron-electron contribution to the electrical resistivity undergo a step-wise increase at the transition point.
		One also finds a significant many-body correction to the $\epsilon$-Fe equation of state, thus clarifying the origin of discrepancies between previous DFT studies and experiment. An electronic topological transition is predicted to be induced in $\epsilon$-Fe by many-electron effects; its experimental signatures are analyzed.
		
		Next section focuses on the geophysically relevant
		pressure-temperature regime of the Earth's inner core (EIC)  corresponding to the extreme pressure of 360~GPa  combined with temperatures up to 6000~K.  The three iron allotropes
		($\alpha$, $\epsilon$  and face-centered-cubic $\gamma$) 
		previously proposed as possible stable phases at such conditions are found to exhibit
		qualitatively different many-electron effects as evidenced by a strongly non-Fermi-liquid
		metallic state of $\alpha$-Fe and an almost perfect Fermi liquid in the case of $\epsilon$-Fe. A recent active
		discussion on the electronic state and transport properties of $\epsilon$-Fe at the EIC conditions is reviewed
		in details. Estimations for the dynamical many-electron contribution to the relative phase stability are presented.  We also discuss the impact of a Ni admixture, which is expected to be present
		in the core matter. We conclude by outlining some limitation of the present DMFT-based
		framework relevant for studies of iron-base systems as well as perspective directions for further
		development. 
		}
	\end{abstract}

\setlength{\parindent}{2em}	
\section*{Introduction }\label{sec:Fe_intoduct}

Iron is a key material for our civilization since the advent of "Iron Age" at about 1000 BC. The technological utility of iron is due to a vast phase space of iron-based alloys, which is the origin of a great variety of steel microstructures obtained with small variations in the composition and an appropriate thermal treatment. In particular,  the rich zoo of steels is composed by three stable phases - the ferrite (body-centered cubic, bcc, $\alpha$) austenite (face-centered cubic, fcc, $\gamma$) and cementite (orthorhombic carbide Fe$_3$C)  - in addition to various metastable phases, for example, the body-centered tetragonal martensite $\alpha'$ (see, e.g., \cite{Bhadeshia_book}).  This multitude of phases observed in iron-based alloys and compounds stems from the complex physics of pure iron, which features three distinct allotropes at the ambient pressure: ground-states bcc $\alpha$-Fe transforms into fcc  $\gamma$-Fe at 1185~K;  the fcc phase subsequently transforms to yet another bcc phase, $\delta$-Fe, at 1667~K. Though $\alpha$ and $\delta$-Fe have the same bcc lattice, their physics is  quite different, with the vibrational entropy believed to be playing the key role  in stabilization of the later \cite{Neuhaus2014}. Iron is a classic itinerant ferromagnet, and the ferromagnetic order is well recognized to be crucial in stabilizing  $\alpha$-Fe \cite{Zener1955}. However, as noted above, the $\alpha$ phase still exists above the Curie temperature of  1044 K. The fcc $\gamma$ phase is thermodynamically stable only at high temperature; this phase is paramagnetic. However, $\gamma$-Fe can be also stabilized in small precipitates in an fcc matrix, e.g., in Cu, down to zero temperature, and at low temperatures it exhibits a complex non-commensurate antiferromagnetic order \cite{Tsunoda1989}. 

\begin{figure}[!t]
	\begin{center}
		\includegraphics[width=0.8\textwidth]{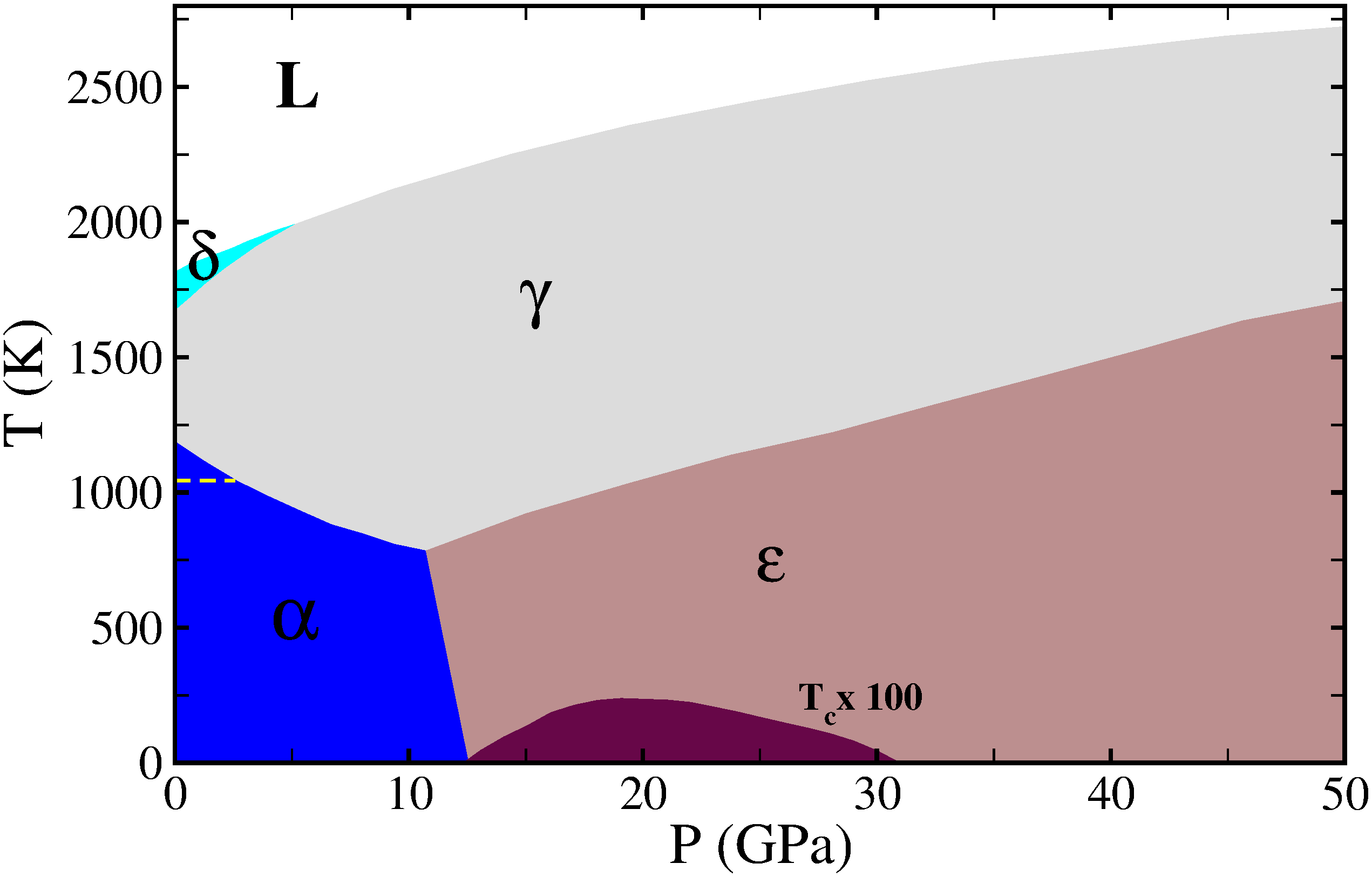}
	\end{center}
	\caption{The pressure-temperature phase diagram of iron in the moderate pressure range up to 50 GPa. The superconducting transition temperature for $\epsilon$-Fe is multiplied by 100. The yellow horizontal dashed line in the $\alpha$-Fe region indicates its ferromagnetic T$_c$. }
	\label{fig:iron_phase_diag}
\end{figure}

Under applied pressure above 10~GPa $\alpha$-Fe transforms into another allotrope, hexagonal close-packed (hcp)  $\epsilon$-Fe~\cite{Bancroft1956,Jamieson1962}. This phase is found to be stable at room temperature up to the highest pressure reached to date~\cite{Mao1990}; {\it ab initio} density-functional-theory (DFT) calculations predict iron to remain in the  $\epsilon$ phase up to a pressure of the order of 10~TPa \cite{Stixrude2012}. Experimental studies of $\epsilon$-Fe reveal a superconducting dome  in the range of pressure from 10 to 30~GPa with the maximum value of superconducting $T_c$ of about 2~K \cite{Shimizu2000}; this superconductivity is likely of non-conventional nature and mediated by spin-fluctuations \cite{Mazin2002}. No magnetic order has been detected in  $\epsilon$-Fe down to temperatures as low as 8~K \cite{Cort1982,Papandrew2006}. A puzzling non-Fermi-liquid (nFL) temperature scaling $\propto T^{5/3}$ of the low-temperature resistivity of $\epsilon$-Fe was also reported \cite{Holmes2004,Yadav2013}.

This rich phase diagram (Fig.~\ref{fig:iron_phase_diag}) with several allotropes exhibiting various magnetic orders, a non-conventional superconductivity as well as instances of a nFL behavior in the $\epsilon$-phase  hint at a complex many-electron physics of iron metal. Before concentrating on the main subject of this review - correlation effects in this metal at high pressure conditions - let us summarize the current picture for the role of these effects at the ambient pressure, for which theoretical predictions can be more easily compared to various experimental probes.

Many-electron effects in iron are expected to arise due to the on-site Coulomb repulsion between rather localized 3$d$ states hybridized with itinerant 4$s$ bands. The typical  width $W$ of the iron 3$d$ band is in the range of 5 to 6~eV for the ambient pressure; the estimated value of the local Coulomb interaction parameter $U$ (Slater $F^0$) is in the range from 2.3 to 6~eV, in accordance with constrained local-density  approximation \cite{Anisimov1991,Cococcioni2005,Belozerov2014} and constrained random-phase approximation \cite{Miyake2008,Miyake2009} calculations (see  Appendix~\ref{dmft:overview} for a short overview of these methods). In spite of a large spread in the theoretical estimates of $U$, one may conclude that the ratio $U/W$ in Fe is less than or equal to 1. Taking into account only the effect of $U \le W$ one would expect rather weak electronic correlation effects in a multiband system away from half-filling \cite{Han1998}. Indeed, the strength of electronic correlations in iron is found to be  much  more sensitive to Hund's coupling $J_H$, which value is in the range of 0.85 to 1~eV. In this respect the physics of iron is close to that of "Hund's metals" \cite{deMedici2011,Haule2009,Georges2013}, in which the strength of correlations away from half-filling is determined mainly by $J_H$. In particular, model studies point out to a key role of $J_H$ in stabilizing the ferromagnetic phase in multiband systems away from half-filling \cite{Fresard1997}. Another important aspect is the interplay between the local Coulomb interaction, characterized by large $J_H$, and crystal-field splitting of Fe 3$d$ states. This interplay is particularly striking in the bcc $\alpha$ phase, where the partial $e_g$ density-of-states (DOS) features a large peak pinned at the Fermi level due to a van Hove singularity  \cite{Maglic1973,Irkhin1993}. Correspondingly, this high DOS at the Fermi level in nonmagnetic $\alpha$-Fe explains its tendency towards the ferromagnetism in accordance with the Stoner criterion. 

The Stoner ferromagnetism of $\alpha$-Fe is well captured by density functional theory (DFT) calculations in conjunction with the local spin-density approximation (LSDA) exchange-correlation functional predicting the theoretical ordered moment of 2.2 $\mu_B$ that agrees well with experiment. Though DFT-LSDA incorrectly predicts $\gamma$-Fe to be the ground states~\cite{Wang1985}, this error is corrected by semi-local exchange-correlation potentials like the generalized gradient approximation (GGA) \cite{Singh1991,Almador1992}. However,  the existence of paramagnetic bcc phase is a significant challenge for density functional theory. Direct DFT calculations predict too small volume and too high bulk modulus for non-magnetic  $\alpha$-Fe; moreover, this  non-magnetic  phase is mechanically and dynamically unstable within DFT \cite{Hsueh2002,Leonov2012}, in clear disagreement with experiment. DFT calculations predict paramagnetic $\gamma$-phase to be dynamically unstable as well \cite{Leonov2012}. A number of methods has been developed in the DFT framework to remedy its deficiency in describing paramagnetic phases. Several such techniques were subsequently applied to iron, like the disordered local moments (DLM) method \cite{Gyorffy1985,Okatov2009,Zhang2011,Khmelevskyi2018}, the spin-statistical-averaging method of K\"ormann {\it et al.}~\cite{Kormann2008}, or the  spin-wave approach of Ruban and Razumovskiy~\cite{Ruban2012},  for a recent review of those techniques see, e.g., Ref.~\cite{Abrikosov2016}.  The spin-disorder contribution to the resistivity of iron at ambient and extreme conditions has been also evaluated using such DFT-based methods \cite{Kudrnovsky2012,Liu2015,Drchal2017,Drchal2019}. However, these techniques represent the paramagnetic state by a certain combination of systems with  static local moments
and are typically useful to describe the thermodynamics of local-moment paramagnets, but not to capture their spectral 
 properties. And even for ferromagnetic $\alpha$-Fe the DFT electronic structure is only in a rough qualitative agreement with experimental photoemission spectra, missing, in particular, the observed quasiparticle renormalization  of the 3$d$ bands by 40-50\% and a lifetime broadening of quasiparticle states \cite{Schafer2005} . 

This inability of pure DFT to fully capture the physics of iron at ambient condition, in particular, of high-temperature paramagnetic $\alpha$-Fe as well as the $\gamma$ and $\delta$ phases has prompted a number of {\it ab initio} studies of this system employing a combination of DFT with a dynamical mean-field theory (DMFT) treatment of the narrow 3$d$ iron band. 

In particular, Leonov and coworkers applied this DFT+DMFT approach in conjunction with a quantum Monte Carlo impurity solver to obtain total energies and phonon dispersions in the paramagnetic $\alpha$ and $\gamma$ phases \cite{Leonov2011,Leonov2012}. Their calculations predict  dynamically and thermodynamically stable paramagnetic $\alpha$-Fe in the range of temperatures from $T_c$ to  1.3$T_c$, in qualitative agreement with experimental phase diagram. Leonov {\it et al.} have subsequently extended their phonon-dispersion calculations of  the bcc phase to the temperature range of $\delta$-Fe~\cite{Leonov2014} finding it dynamically unstable in the harmonic approximation, this result was very recently challenged by another DFT+DMFT study~\cite{Han2017}. Theoretical DFT+DMFT calculations of the one-electron spectra of iron \cite{Katsnelson1999,Lichtenstein2001,Minar2005,Grechnev2007,Sanchez-Barriga2009,Hausoel2017,Han2017} have been mostly limited to the ferromagnetic $\alpha$ phase, for which experimental angular-resolved photoemission (ARPES) spectra are available~\cite{Schafer2005,Sanchez-Barriga2009}. Refs.~\cite{Katanin2010} and \cite{Igoshev2013}  also studied the one-electron spectral function and magnetic susceptibilities of the paramagnetic $\alpha$ and $\gamma$ phases. S\'anchez-Barriga {\it et al.}~\cite{Sanchez-Barriga2009} concluded that a purely-local single-site DMFT self-energy is not sufficient to obtain a quantitative agreement  between the theoretical $\vk$-resolved spectral function and experimental ARPES spectra, though they employed an approximate treatment of the DMFT quantum impurity problem. The most recent DFT+DMFT studies~\cite{Hausoel2017,Han2017} employing a numerically-exact quantum Monte Carlo approach \cite{Gull2011} reach a reasonable quantitative agreement with ARPES, though discrepancies for some high-symmetry directions are  still present.   The same level of agreement was reached by including both local and non-local many-electron effects within a weak-coupling quasiparticle GW approach~\cite{Sponza2017}. Hence, a combination of non-perturbative treatment of the on-site correlations with a weak-coupling approach to non-local ones (see, e.g., Refs.~\cite{Biermann2003,Tomczak2017}) is probably necessary to fully account for the one-electron spectra of ferromagnetic $\alpha$-Fe.

Correlation effects in iron under moderate and high pressure have been less studied with DFT+DMFT until recently. The present review focuses on this topic, in particular, on new theoretical results obtained during the last 5 years. 

First, we consider the hcp $\epsilon$ phase, whose  ground-state and transport properties  in the moderate pressure range of 10 to 60 GPa are puzzling, as shortly described above. 
The subsequent section discusses with properties of the $\alpha$, $\gamma$ and $\epsilon$ iron and iron-nickel alloy at the volume of 7.05 \AA/atom and at temperatures up to 6000~K. These density and temperature are expected for the inner core of Earth, hence,  the phase stability and transport properties of iron at such conditions are of a particular relevance to the geophysics. In spite of high density, which is expected to diminish the relative importance of the electron-electron repulsion, we still find a rather significant impact of the local interaction between  3$d$ electrons on the electronic structure, phase stability as well as on magnetic and transport properties. 

We supplement this review with Appendix A briefly presenting the DFT+DMFT methodology, which has been employed to calculate the key quantities of the many-electron theory referred in the main text, like the one-electron Green's function and self-energy.  The DFT+DMFT method has not yet become a standard tool in {\it ab initio} calculations of transition metals and their alloys. This technique has been much more intensively applied to "traditional" strongly correlated materials, like transition-metal oxides and heavy-fermion systems. Hence, the Appendix will hopefully appeal to  theoretically oriented readers working in the field of transition-metal alloys  and interested in beyond-DFT approaches to electronic correlations in these systems. Some familiarity with DFT and basics of many-body quantum theory (like the second quantization and Green's function formalisms)  is thus assumed.  Appendix A is not necessary to understand the main text; the  readers that are mainly interested in experimentally accessible results and not in the methodology may skip it.  

\section{$\epsilon$-Fe under moderate pressure: equation of state, resistivity and electronic topological transitions}\label{sec:epsFe}

As noted above, DFT  successfully captures the magnetic state $\alpha$-Fe; DFT calculations also predict the ground-state properties of this phase in good agreement with experiment. 
In contrast, the same theory fails to account even for basic ground-state properties of pressure-stabilized $\epsilon$-Fe. Within the local spin-density and generalized gradient approximations for the exchange-correlation potential it predicts a rather strong antiferromagnetism (AFM), with the iron moment of about 1.5~$\mu_B$ at the volume of 73 (a.u.)$^3$/atom, corresponding to that of the $\epsilon$-phase at the $\alpha \to \epsilon$ transition point \cite{Steinle-Neumann1999,Steinle-Neumann2004,Steinle-Neumann_2004_2}. However, no hyperfine magnetic splitting has been observed in $\epsilon$-Fe by M\"ossbauer spectroscopy down to 8~K \cite{Cort1982,Papandrew2006} questioning the existence of any static magnetic order in this phase. X-ray emission spectroscopy of Refs.~\cite{Monza2011,Lebert2019} detected a magnetic signal, but it is not clear whether its origin is a static order or rapidly fluctuating local moments. The same very recent work \cite{Lebert2019}  proposed a quasi-2d  order of alternating AFM and  "magnetically dead" layers  for $\epsilon$-Fe but their neutron diffraction measurements failed to detect any magnetic order in this phase down to 1.8~K at 20~GPa.  The presence of a high-frequency satellite  of the $E_{2g}$ Raman mode in $\epsilon$-Fe \cite{Merkel2000} was initially ascribed to a splitting of this mode in the AFM-ordered phase \cite{Steinle-Neumann2004}; the satellite peak is, however, found to disappear at low temperatures, i.~e., where the AFM state is supposed to be stable \cite{Goncharov2003}.     

Hence, at present there is no convincing evidence in favor of any ordered magnetic state in $\epsilon$-Fe. If the nonmagnetic ground state is imposed, DFT total energy calculations predict an equation of state that drastically disagrees with experiment. The bulk modulus is overestimated by more than 50\%, and the equilibrium volume is underestimated by 10\% compared to the experimental values \cite{Steinle-Neumann1999}. Another puzzling experimental observation is a large enhancement in the resistivity across the $\alpha$-$\epsilon$ transition. The room temperature total resistivity of $\epsilon$-Fe is twice as large as that of the $\alpha$ phase \cite{Holmes2004}. The electron-phonon-scattering contribution to resistivity calculated within GGA is in excellent agreement with the experimental total resistivity for the $\alpha$ phase \cite{Sha2011}, however, these calculations predict virtually no change in the resistivity across the transition to antiferromagnetic hcp-Fe. Moreover, the measured temperature dependence of resistivity in $\epsilon$-Fe exhibits $\sim T^{5/3}$ temperature dependence \cite{Holmes2004,Yadav2013}, the $n=5/3$ exponent is characteristic of weakly ferromagnetic metals, in disagreement with the AFM tendency predicted by DFT.  All these discrepancies between DFT calculations and experiment point out to a possible important role of dynamic correlations in $\epsilon$-Fe. Sola {\it et al.}~\cite{Sola2009} performed diffusion quantum Monte Carlo  (DQMC) calculations of the $\epsilon$-Fe equation of state  thus including many-electron effects beyond DFT. Their resulting equation of state is very similar to the DFT one and strongly deviates from experiment at low pressures, where correlation effects are most important. This may be due to the fixed-node approximation employed in DQMC with the nodal surface fixed by a trial wave function obtained within DFT.  

\begin{figure}[!t]
	\begin{center}
		\includegraphics[width=0.8\textwidth]{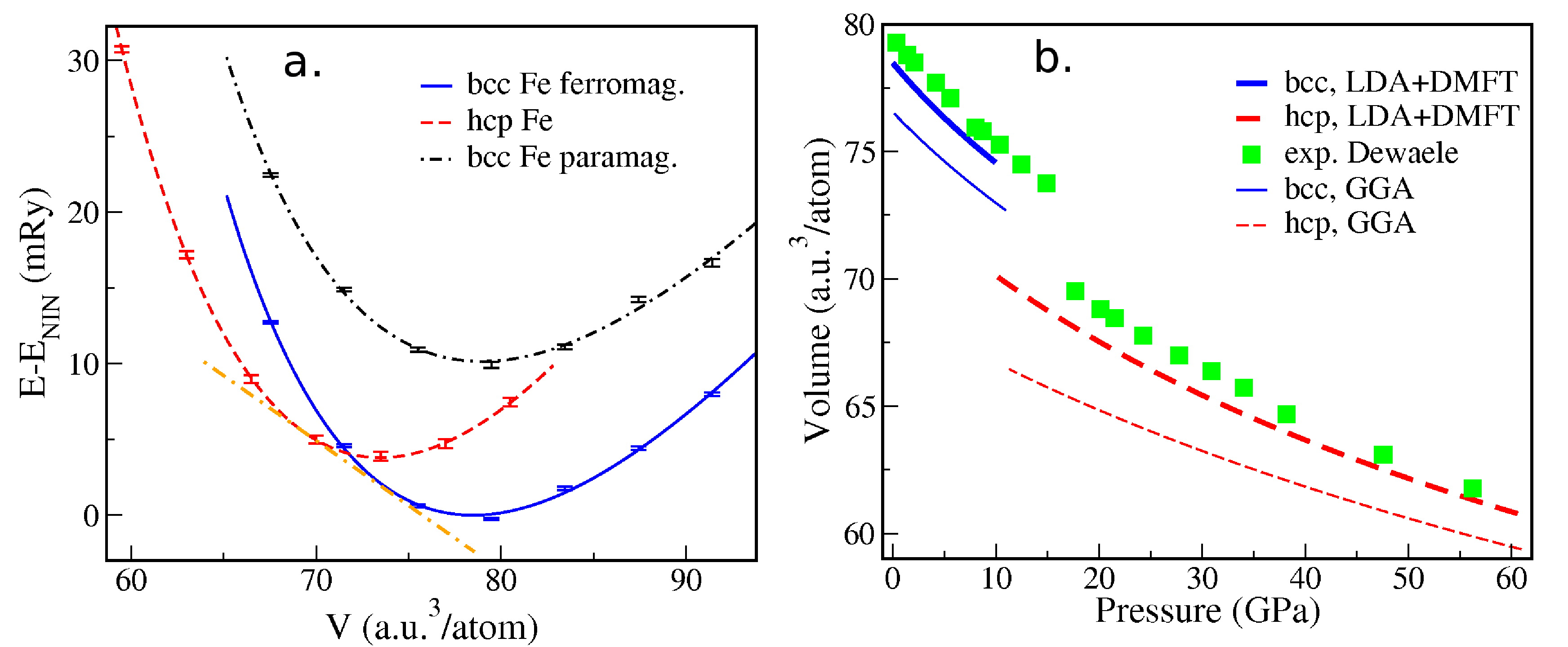}
	\end{center}
	\caption{a). DFT+DMFT total energy vs. volume per atom for $\alpha$ (ferromagnetic, solid blue line, and paramagnetic, dot-dashed black line) and $\epsilon$ (dashed red line) Fe. The error bars are the CT-QMC method stochastic error. The orange long dash-dotted straight line indicates the common tangent construction for the $\alpha-\epsilon$ transition. b). Equations of states (EOS) for ferromagnetic $\alpha$ (low pressure) and paramagnetic $\epsilon$ (high pressure) Fe. Theoretical results are obtained by fitting  the DFT+DMFT (thick line) and GGA (thin line) total energies, respectively,  using the Birch-Murnaghan EOS. The experimental EOS of iron shown by green filled squares is from  \cite{Dewaelle2006}. Adapted from Ref.~\cite{Pourovskii2014}.}
	\label{fig:hcpFe_eq_of_state}
\end{figure}

The evolution of electronic correlations across the $\alpha \to \epsilon$ transition as well as its impact on the equation of state and electrical resistivity were studied by Pourovskii {\it et al.}~\cite{Pourovskii2014} using a self-consistent DFT+DMFT approach; the quantum impurity problem was solved using the hybridization-expansion CT-QMC method briefly introduced in Appendix Sec.~\ref{sec:QIP}. In order to achieve the necessary accuracy with a manageable computational cost the so-called  non-density-density terms in the Coulomb vertex were neglected, see Appendix~\ref{chap:appendix_FeDensDens}. This local Coulomb interaction vertex between Fe 3$d$ states was parametrized by $U=F^0=$4.3~eV and $J_H$=1~eV. These values of the interaction parameters were chosen on the basis of the previous cRPA calculations for iron by \cite{Miyake2009}; their value of  $U=$3.4~eV for $\alpha$-Fe was increased by about 25\% to effectively account for the high-frequency tails of the Coulomb vertex \cite{Casula2012}. The value of $J_H$ was fixed at the top of the accepted range of 0.85 to 1.0~eV  to reproduce the value of magnetic moment in $\alpha$-Fe at the ambient conditions.

Overall, DFT+DMFT total energy calculations of Ref.~\cite{Pourovskii2014} provide a comprehensive and quantitatively correct picture for the ground-state properties of both phases including their ground-state volumes, bulk moduli as well as the pressure  dependence of the $c/a$ ratio in $\epsilon$-Fe. In particular, they predict a ferromagnetic $\alpha$-Fe ground state and a transition $\alpha \to \epsilon$ phase at 10~GPa, in agreement with experiment (Fig.~\ref{fig:hcpFe_eq_of_state}a). The calculated difference in total energy between the ferromagnetic and paramagnetic states of $\alpha$-Fe is of about 10~mRy (1500~K), in a good correspondence to its experimental Curie temperature of 1043~K. The Birch-Murnaghan equations of states  (EOS) fitted to DFT+DMFT total energies of $\alpha$ and $\epsilon$-Fe   agree well with the corresponding experimental EOS  (Fig.~\ref{fig:hcpFe_eq_of_state}b). One observes a particularly significant improvement for the case of $\epsilon$-Fe, for which the DFT-GGA framework performs quite poorly.   In contrast, the DFT+DMFT corrections to EOS of ferromagnetic $\alpha$-Fe are rather small; as noted in the Introduction, the DFT in conjunction with GGA already describes the ground-state properties of this phase quite well. 

\begin{figure}[!t]
	\begin{center}
		\includegraphics[width=0.7\textwidth]{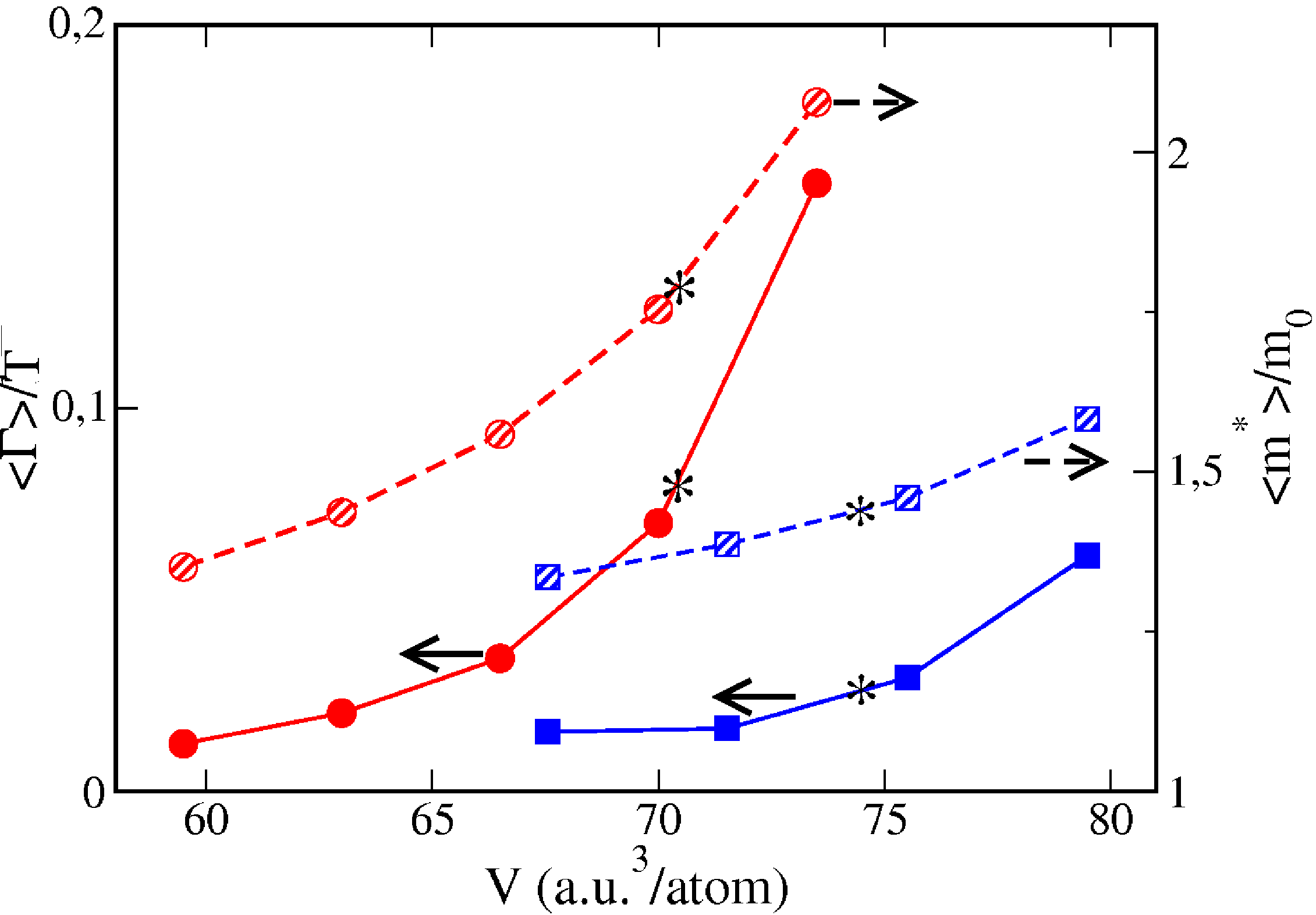}
	\end{center}
	\caption{The ratio of averaged over orbitals inverse quasiparticle lifetime $\langle \Gamma\rangle$ to
		temperature (the left axis) and the analogously averaged mass enhancement
		$\langle m^*\rangle/m_0$ (the right axis) vs. volume per atom. The
		solid lines (filled symbols) and dashed lines (hatched symbols) are
		$\langle\Gamma\rangle/T$ and $\langle m^*\rangle/m_0$,
		respectively. The values for $\alpha$ and $\epsilon$ phases are shown by blue
		squares and red circles, respectively. The black stars indicated
		their corresponding atomic volumes at the transition point. Adapted from Ref.~\cite{Pourovskii2014}.} 
	\label{fig:hcpFe_eff_m}
\end{figure}

As the many-body corrections to the ground-state properties are more significant in $\epsilon$-Fe than in $\alpha$-Fe one can expect stronger dynamic electronic correlations in the former. Indeed, the mass enhancement $m^*_{a}/m_0=Z^{-1}_{a}$ and the inverse quasiparticle lifetime 

\begin{equation}\label{eq:Gamma}
\Gamma_{a} =-Z_{a}\Im\Sigma_{a}(\omega=0),
\end{equation}
where $a$ is the $m$ and spin quantum numbers labeling Fe 3$d$ orbitals, $Z_{a}=\left[1-\left.d\Im\Sigma_{a}(i\omega)/d\omega\right|_{\omega\to
	0}\right]$ is the quasiparticle residue
extracted from the  zero-frequency value of the DMFT self-energy  $\Sigma_{a}$ (see Appendix Sec.~\ref{sec:dmft}) for the orbital $a$,   exhibit a large increase at the $\alpha \to \epsilon$ transition (Fig.~\ref{fig:hcpFe_eff_m}).  This enhancement of dynamic correlations is due to the suppression of the static magnetic order at this transition. In fact, paramagnetic $\alpha$-Fe is a strongly-correlated nFL system, with a particularly large value of $\Gamma$ for localized $e_g$ states \cite{Katanin2010,Pourovskii2013}. In contrast, only a modest Fermi-liquid renormalization of Fe 3$d$ DFT band structure is detected by ARPES for the ferromagnetic  phase \cite{Schafer2005}; their value for the mass enhancement of about 40-50\% agrees reasonably with the DFT+DMFT prediction of 1.6  for $\langle m^*\rangle$ for the ambient conditions (Fig.~\ref{fig:hcpFe_eff_m}). 

A step-wise increase of the inverse quasiparticle lifetime $\Gamma$ at the $\alpha \to \epsilon$ transition point should result in a corresponding step-wise increase of the electron-electron-scattering contribution to the electrical resistivity. Indeed, DFT+DMFT calculations for the transport presented in the same paper \footnote{See Sec.~\ref{subsec:epsFe_transport} for a brief summary of the formalism for transport calculations} predict such a jump with the electron-electron contribution enhanced by a factor of 3, from 0.5~$\mu\Omega\cdot$cm in $\alpha$-Fe to 1.5~$\mu\Omega\cdot$cm in the $\epsilon$ phase. The jump in total resistivity $\rho$ at the transition  observed experimentally \cite{Holmes2004,Yadav2013} features an overall qualitative shape of the resistivity vs. pressure in iron strongly resembling the DFT+DMFT one. However, the experimental jump in $\rho$ at the $\alpha \to \epsilon$ transition for the room temperature is an order of magnitude larger than 1~$\mu\Omega\cdot$cm predicted by our calculations. The present  approach, apparently, misses the main source of this resistivity enhancement. The fact that the resistivity jump is still well resolved at $T=$4~K lends a strong support to its electron-electron-scattering origin. A strongly nFL  behavior of $\epsilon$-Fe in the temperature range from 2 to (at least) 30~K \cite{Holmes2004,Yadav2013}, in conjunction with a non-conventional superconducting state at lower $T$  points out at important intersite correlations, e.g. spin fluctuations, which are neglected by the single-site DFT+DMFT framework. Alternatively, one may suggest that local non-density-density  interaction terms (see Appendix \ref{sec:QIP} and \ref{chap:appendix_FeDensDens}) neglected in Ref.~\cite{Pourovskii2014} have a crucial impact on the low-energy behavior of the self-energy $\Sigma(\omega)$ and, hence, at the transport. This problem is an interesting subject for future work.

\begin{figure}[!t]
	\begin{center}
		\includegraphics[width=0.50\textwidth]{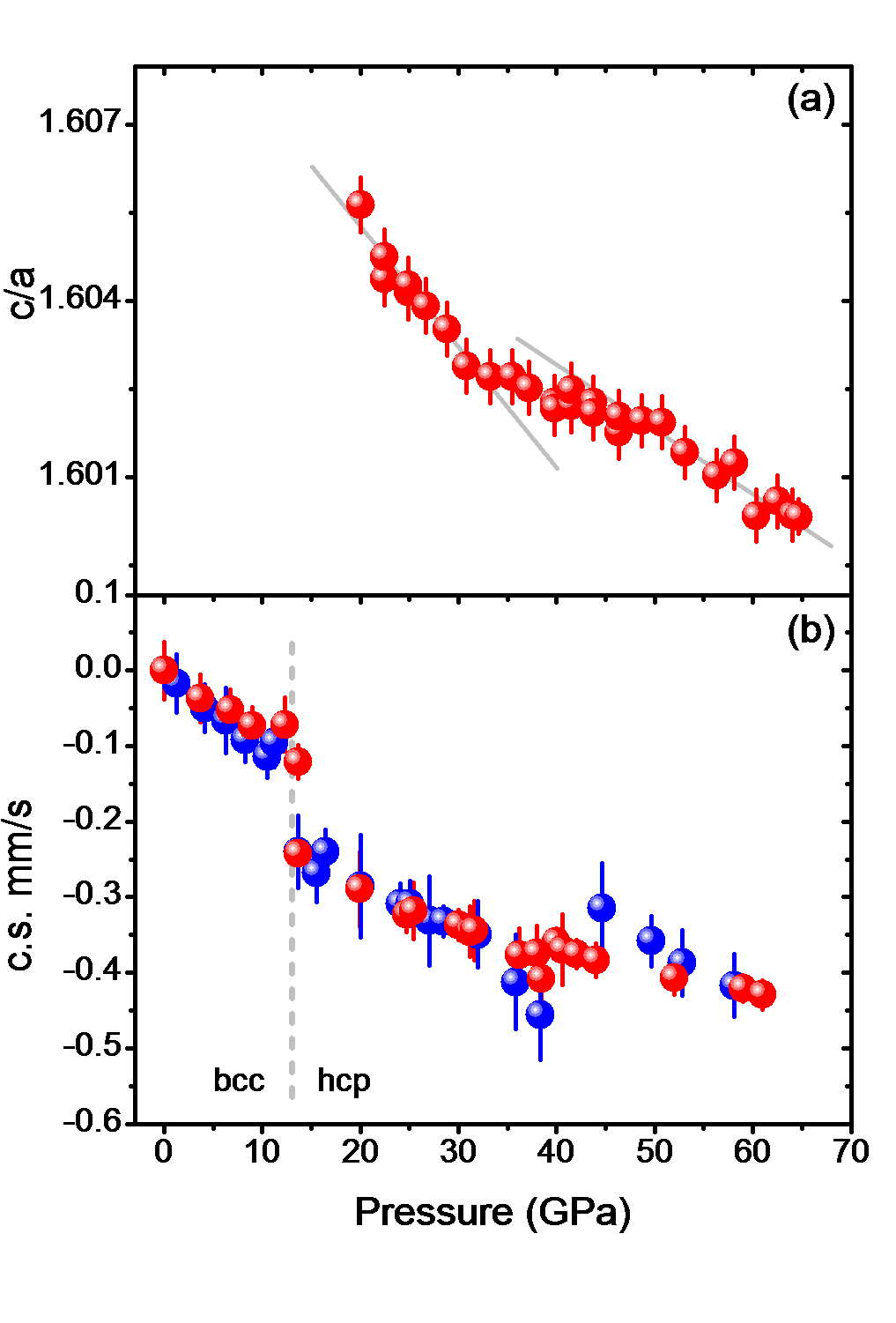}
	\end{center}
	\caption{ Experimental pressure dependence of (a) hcp phase $c/a$ ratio and (b) the M\"ossbauer centre shift based on several experimental datasets for pure iron (red circles) and for Fe$_{0.9}$Ni$_{0.1}$ alloy (blue circles). The centre shift values are given relative to pure $\alpha$ iron. Straight grey lines in (a) are guides for the eye. Adapted from Ref.~\cite{Glazyrin2013}.} 
	\label{fig:hcpFe_ETT_exp}
\end{figure}

No experimental ARPES of $\epsilon$-Fe has been reported to date
as such measurements are not feasible at a high pressure of tens GPa. Glazyrin {\it et al.} \cite{Glazyrin2013} studied the impact of pressure on the electronic structure of the $\epsilon$ phase by measuring a set of quantities readily accessible at high pressure conditions, namely, the Debye sound velocity, M\"ossbauer central shift and hexagonal cell $c/a$ ratio, in pure Fe and in Fe$_{0.9}$Ni$_{0.1}$.  All three quantities are found to exhibit a distinct peculiarity at about 40~GPa. One sees, for example, a clear change of slope in the evolution of $c/a$ vs. $P$  as well as a peculiarity in the M\"ossbauer central shift at this pressure, which is especially pronounced in the case of Fe$_{0.9}$Ni$_{0.1}$ (Fig.~\ref{fig:hcpFe_ETT_exp}). As discussed by  Ref.~\cite{Glazyrin2013}  peculiarities simultaneously appearing in all three quantities  can be qualitatively explained by an electronic topological transition (ETT) due to the appearance of new Fermi-surface hole pockets at a given pressure \cite{Vaks1991,Novikov1999,Katsnelson2000}. The resulting peculiarities in these quantities are proportional to the change of DOS at the Fermi level, $\delta N(E_F)$, due to the ETT.

\begin{figure}[!t]
	\begin{center}
		\includegraphics[width=1.0\textwidth]{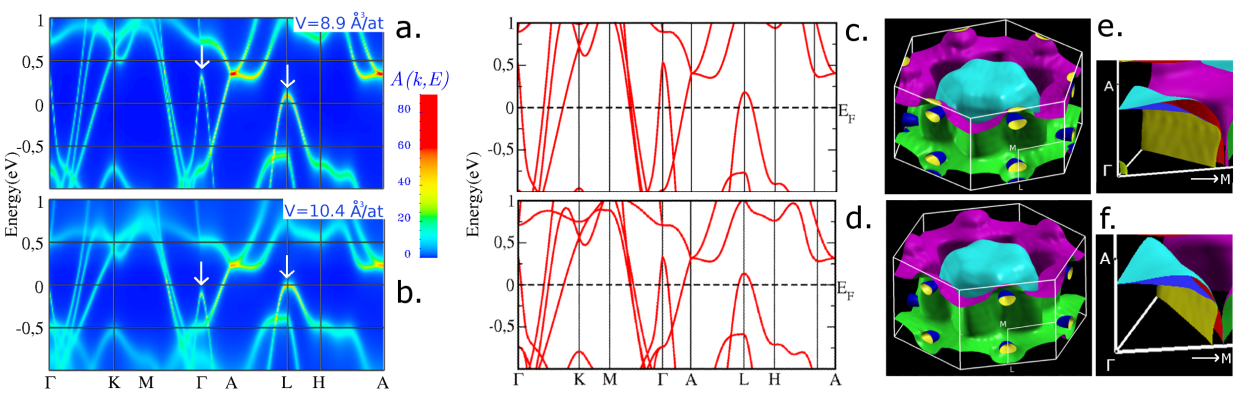}
	\end{center}
	\caption{The DFT+DMFT $\vk$-resolved spectral function $A(\vk,\omega)$ ( in units of V$_{at}$/eV, where V$_{at}$ is the volume per atom) of $\epsilon$-Fe at volumes V$_{at}$ of 8.9 \AA$^3$/atom (a) and 10.4 \AA$^3$/atom (b) corresponding to pressures of 69 and 15.4 GPa, respectively. 		
		The energy zero is taken at the Fermi level. The hole-like bands at the $\Gamma$ and L points at volume 8.9 \AA$^3$/atom (indicated by the white arrows) are below $E_F$ at V=10.4 \AA$^3$/atom. The corresponding DFT band structures are shown in c and d, respectively. The corresponding DFT+DMFT Fermi surfaces for two volumes are shown in e and f, respectively. Adapted from Ref.~\cite{Glazyrin2013}.} 
	\label{fig:hcpFe_ETT_theory}
\end{figure}

In order to precisely identify the ETT at the origin of observed peculiarities Glazyrin {\it et al.}~\cite{Glazyrin2013}  calculated  the DFT+DMFT $\vk$-resolved spectral function $A(\vk,\omega)=-\frac{1}{\pi}\Im G(\vk,\omega+i\delta)$) from  the analytically-continued lattice Green's function (GF), eq. \ref{eq:dyson} in Appendix~\ref{sec:dmft}, as a function of volume.  $A(\vk,\omega)$ obtained by DFT+DMFT clearly features the emergence of new hole pockets at the $\Gamma$ and $L$ high symmetry point (Fig.~\ref{fig:hcpFe_ETT_theory}a and \ref{fig:hcpFe_ETT_theory}b). The corresponding critical pressure for the ETT is found to be in the range of 40-80 GPa, depending on the chosen value of $U$. In contrast, the DFT band structure features those hole pockets (Fig.~\ref{fig:hcpFe_ETT_theory}d) already at  10.4 \AA$^3$/at, which is the atomic volume of $\epsilon$-Fe at the $\alpha \to \epsilon$ transition. Hence, DFT does not predict any ETT to occur in the $\epsilon$ phase in its experimental range of existence.

This significant shift of ETT to lower volumes/higher pressures in DFT+DMFT compared to pure DFT are mainly due to many-electron corrections to the overall position of the valence $d$ bands with respect to the $s$ ones, leading to a relative shift of states with a significant $s$ contribution with respect to the rest. 
A similar significant impact of many-body corrections was recently predicted even for such weakly correlated system as the osmium metal by Feng {\it et al.}~\cite{Feng2017}. They found the transition pressures for a series of ETTs to be in a better agreement with experiment when DMFT corrections were included. One may notice, however, that the relative shift of "correlated" $d$ vs. "uncorrelated" $s$ states is sensitive to the choice of the double-counting (DC )correction (see Appendix~\ref{dmft:overview}). Both Refs.~\cite{Pourovskii2014} and \cite{Feng2017} employ the "around mean-field" form of DC, which is believed to be appropriate for such relatively itinerant systems. 

On the experimental side,  Dewaele and Garbarino~\cite{Dewaele2017} have very recently reported new  measurements of the equation of state and $c/a$ ratio of $\epsilon$-Fe. The experimental equation of state is  found to be in good agreement with calculations of Ref.~\cite{Pourovskii2014}. Although no sign of peculiarity was observed in the $c/a$ ratio by Ref.~\cite{Dewaele2017}, one may notice that the scatter of their points is significantly larger than that of  Glazyrin {\it et al.}~\cite{Glazyrin2013}.

\section{Many-electron effects in iron and iron-nickel alloy at the Earth's inner core conditions}\label{sec:Fe_in_core}

After the previous discussion focused on the moderate pressure range of several tens GPa, we move on to a much more "extreme" regime of pressures above 300 GPa and temperatures of several thousands Kelvins. This region in the pressure-temperature phase diagram of iron is believed to be directly relevant to the structure and dynamics of the deep interior of our planet. 

The wealth of available data on seismic wave propagation, planetary density and gravitational field, abundance of elements in the Solar system lends strong support to the hypothesis of iron being one of principal component of Earth and Earth-like planets~\cite{Birch1952,Dziewonski1981,Lowrie2007}. In particular, the solid inner and liquid outer cores of Earth are believed to consist mainly of iron. The measured Earth interior density profile as well as data on the meteorite composition favor a picture of a solid Earth's inner core (EIC) composed of iron alloyed with about 10\% of nickel and non-negligible quantities of light elements like Si, S, or O.  Inside the EIC the matter is subjected to pressure $P$ in the range of 330 to 360 GPa at temperature $T$ of about 6000~K, though the relevant range of $T$ for the inner core is still actively under debate. The temperature of solid phase inside EIC is close to its melting point. The experimental melting curves  are not to date yet reaching the EIC pressure and need to be extrapolated resulting in the range of about 4800-6200~K for the EIC boundary~\cite{Boehler1993,Shen1998,Nguyen2004,Jackson2013,Anzellini2013,Sinmyo2019}; DFT-based molecular dynamics \cite{Alfe1999,Laio2000,Belonoshko3638} and DQMC calculations \cite{Sola2009_1} produce similar, though somewhat higher estimates for the temperature at the EIC boundary.  The phase stability and properties of solid iron and iron-rich alloys at such extreme conditions are of high importance for the geophysics as they represent a key input to geophysical models of Earth's core dynamics and its evolution. In particular, the interpretation  of  seismic data is largely based on the assumed phase diagram for relevant iron-rich alloys at  the core's conditions~\cite{Tkalcic2015}. The models of core evolution in time are constrained by the accepted range of  values for  the thermal and electrical conductivities~\cite{Buffett2012,Pozzo2012}. Therefore, significant research efforts, both experimental and theoretical, are focused on reliably  determining the nature of Fe phases stable in the relevant ($P$,$T$) range and their physical properties.

Iron and its alloys at extreme conditions were initially studied experimentally using the dynamical shock-wave compression~\cite{Brown1986} and, more recently, also with the static heated diamond anvil cell method. As noted in the previous section, these studies have established the stability of $\epsilon$-Fe up to the pressure range of EIC at the room temperature~\cite{Mao1990}. The situation is less clear for the high-T region, where some recent experiments~\cite{Nguyen2004,Tateno2010,Tateno2012,Anzellini2013,Ping2013} found the $\epsilon$-phase in the relevant pressure range up to the EIC temperatures, while other studies \cite{Dubrovinsky2007,Hrubiak2018} observed bcc $\alpha$-Fe to emerge at high temperatures approaching the melting point. Tateno {\it et al.}~\cite{Tateno2010} claimed to reach the EIC conditions in their anvil-cell experiments and observed only the $\epsilon$ phase in the studied range of $P$ from 100~GPa to the highest pressure of 377 GPa and $T$ from 2000 to 5700~K. However, their interpretation of the data was subsequently disputed by Dubrovinsky {\it et al.}~\cite{Dubrovinsky2011}, who suggested that the EIC temperature was not in fact reached by Tateno {\it et al.}~\cite{Tateno2010}. Overall, currently there is no experimental  consensus regarding the stable phase of Fe at EIC conditions.

The theory input is particularly valuable in such situation, hence, a number of DFT based simulations of Fe and its alloy has been published in the last two decades. 
These studies treated lattice vibrations in the quasi-harmonic approximation~\cite{Mikhaylushkin2007,Stixrude2012} or with the full {\it ab initio } molecular dynamics approach~\cite{Vocaldo2003,Godwal2015,Belonoshko2017}. The results of these calculations are also inconclusive, with all three known phase of iron predicted to be stable at EIC conditions by different authors. The difference in DFT free energy between those phases is found to be decreasing with increasing temperature and pressure. Thus the relative stability becomes sensitive to small differences in the calculational setup like the size of simulation supercell or the density of $\vk$-mesh employed in the Brillouin zone integration \cite{Godwal2015,Belonoshko2017}. 
In particular, the non-magnetic $\alpha$ phase dynamically unstable at low temperature is claimed by Belonoshko {\it et al.}~\cite{Belonoshko2017} to be stabilized by an unconventional high-T diffusion mechanism; in contrast, Godwal {\it et al.}~\cite{Godwal2015} found $\alpha$-Fe to be dynamically unstable at the EIC conditions. The free-energy difference between $\gamma$ and $\epsilon$-Fe becomes extremely small close to the melting temperature in accordance with Ref.~\cite{Mikhaylushkin2007}, who predicted $\gamma$-Fe to be stable at the EIC conditions, while Stixrude~\cite{Stixrude2012} found the $\epsilon$-phase to be more stable. In all these {\it ab initio} simulations the standard DFT framework in the conjunction with the local-density approximation (LDA) or GGA exchange-correlation potentials was employed thus neglecting dynamical correlation effects. This approximation is usually justified (see, e.~g., Ref.~\cite{Stixrude2012}) by the fact that the local Coulomb repulsion $U$ between iron 3$d$ states is smaller than the effective 3$d$ bandwidth, especially at high pressure.
Though  this statement is correct even at the ambient pressure, this does not mean that correlation effects in iron are negligible. As noted in the previous section, the strength of local many-electron effects in iron is much more sensitive to the Hund's rule coupling $J_H$, which is expected to be quite insensitive to pressure. High temperature stabilizing high-entropy states  may strengthen the tendency towards a nFL behavior or the formation of local magneitc moments. Hence, the role of many-electron effects needs to be evaluated with explicit calculations. 

This problem was addressed  by  DFT+DMFT calculations~\cite{Pourovskii2013} for the all three phases, $\alpha$, $\gamma$ and $\epsilon$, for the volume of 7.05 \AA$^3$/atom, corresponding to the density of EIC, and for temperatures up to 5800~K by
employing the same self-consistent in the charge density full-potential DFT+DMFT framework 
as in the studies of $\epsilon$-Fe described  in the previous section.  
This work evaluated the impact of many-electron effects on the electronic structure, magnetic susceptibility and relative stability of the three Fe phases.  All DFT+DMFT calculations were done for the corresponding perfect fixed lattices. Lattice vibrations play a paramount role at the extreme temperatures inside the EIC, but including their effect within a kind  of DFT+DMFT-based molecular dynamics is prohibitively costly at present. The fixed-lattice calculations of Ref.~\cite{Pourovskii2013}, however, allowed evaluating the structural dependence of correlation effects, assessing (though quite roughly) their impact on the electronic free energy "landscape" in the structural coordinates. Subsequent works~\cite{Vekilova2015,Hausoel2017} carried out similar calculations for Fe-rich FeNi alloys in order to assess the impact of Ni substitution on many-electron effects. A later study of Ref.~\cite{Pourovskii2017} concentrated on the $\epsilon$-phase evaluating its electronic state as well as electrical and thermal conductivities. The results obtained in these works for the electronic structure, magnetism, thermodynamic stability and transport are reviewed below.


\subsection{Electronic structure and magnetic susceptibility of iron}\label{subsec:Fe_in_core_elstruct}

\begin{figure}[!b]
	\begin{center}
		\includegraphics[width=0.55\textwidth]{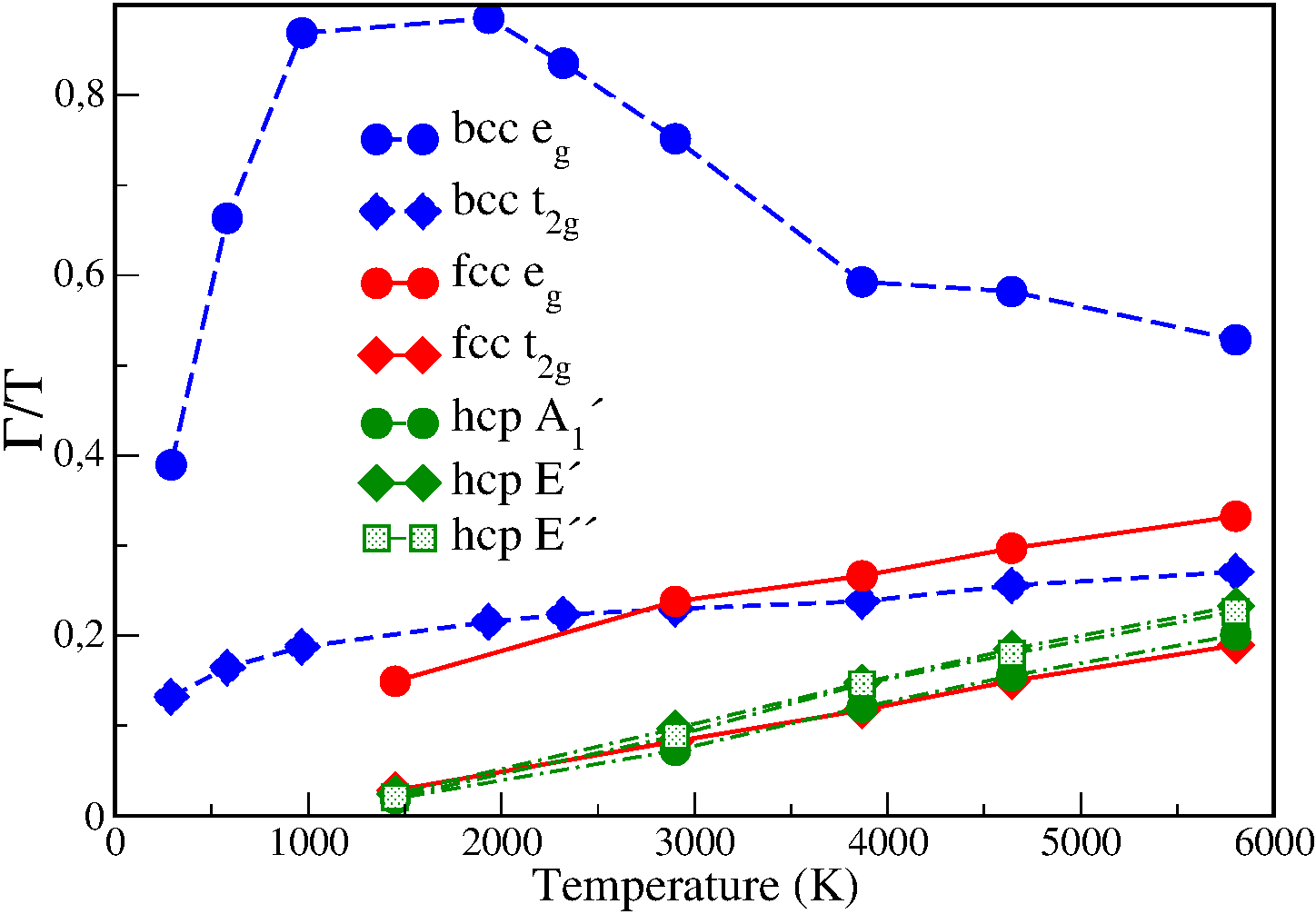}
	\end{center}
	\caption{ The ratio of the inverse quasiparticle lifetime $\Gamma$ to temperature $T$ vs. $T$.
		The solid red, dashed blue and dash-dotted green curves correspond to 3d states in fcc, bcc, and hcp Fe,
		respectively. They are split by the crystal field into $t_{2g}$ (diamonds) and $e_g$ (circles) representations in the cubic
		(bcc and fcc) phases, and two doubly-degenerate ($E'$ and $E''$, shown by diamonds and squares, respectively) and 
		one singlet ($A_1'$, circles) representations in the
		hcp phase, respectively . A non-linear behavior of $\Gamma$/T for bcc Fe $e_g$ states  is clearly seen. Adapted from Ref.~\cite{Pourovskii2013}.} 
	\label{fig:Gamma_Fe_all_phases}
\end{figure}

The ratio  $\Gamma/T$ (see eq. \ref{eq:Gamma})   calculated at the EIC atomic volume  as a function of $T$ in Ref.~\cite{Pourovskii2013} is shown in Fig.~\ref{fig:Gamma_Fe_all_phases} for all relevant irreducible representations of the three phases. One may readily notice a qualitative difference in the behavior of $\Gamma$ between these phases. The  temperature scaling $\Gamma/T \propto T$ expected in the case of a good FL is clearly observed for the $\epsilon$-phase. In contrast,  $\Gamma/T$ for the bcc iron $e_g$ states
features a linear and steep rise for $T < $ 1000~K and then behaves  non-linearly, indicating a
non-coherent nature of those states at high temperatures. The bcc Fe $t_{2g}$ and fcc Fe $e_g$ electrons
are in an intermediate situation with some noticeable deviations from the FL behavior.

\begin{figure}[!t]
	\begin{center}
		\includegraphics[width=0.7\textwidth]{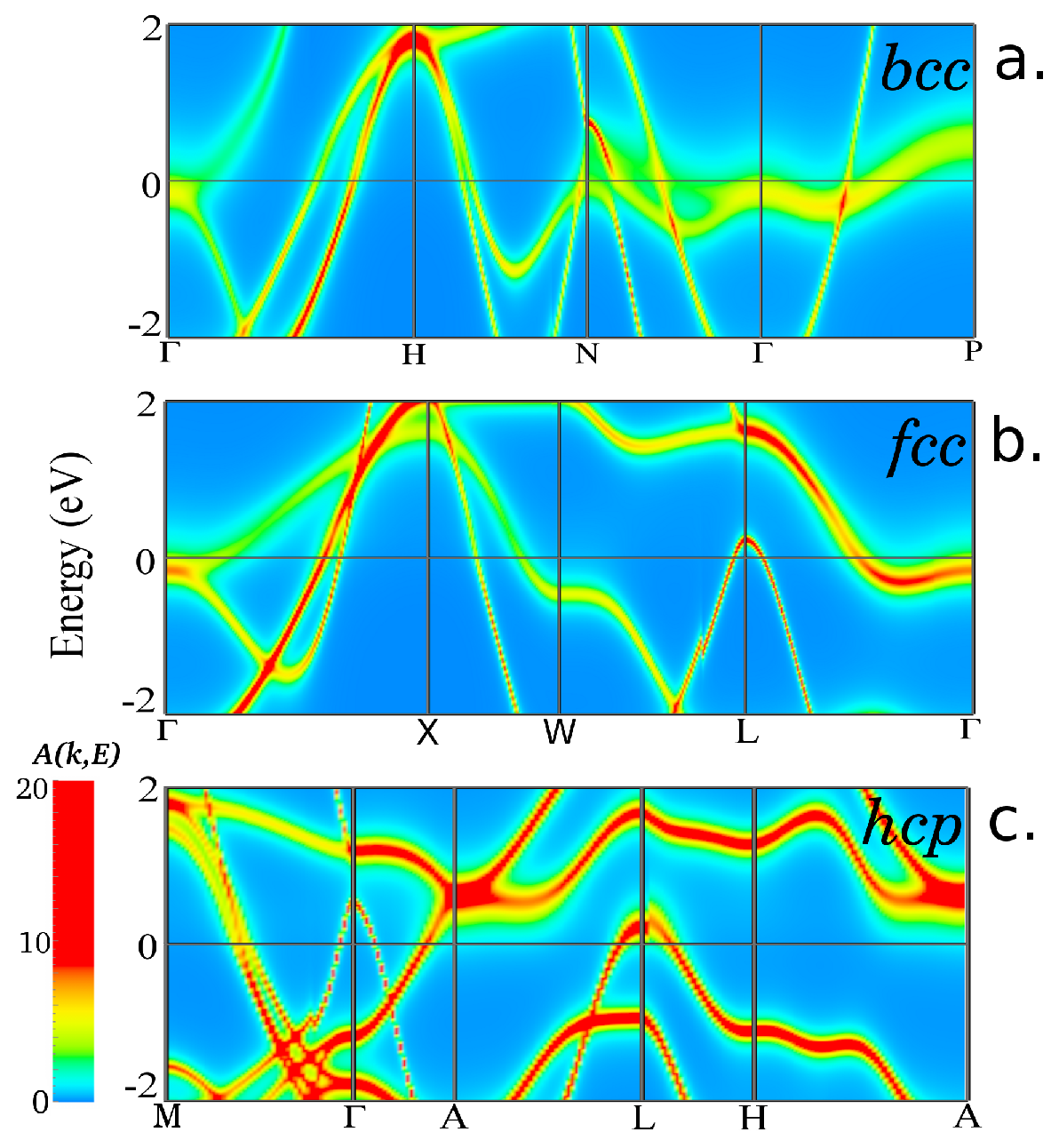}
	\end{center}
	\caption{ The DFT+DMFT $\vk$-resolved spectral function $A(\vk,\omega)$ (in $V_{atom}$/eV) for bcc (a), fcc (b), and
		hcp (c) Fe at volume $V_{atom}=$7.05 \AA$^3$/atom and temperature 5800~K. A non-quasiparticle $e_g$ band is seen in the vicinity of the Fermy energy along the $N-\Gamma-P$ path in (a). Adapted from Ref.~\cite{Pourovskii2013}.} 
	\label{fig:Akw_Fe_all_phases}
\end{figure}

The same conclusions can be drawn from the $\vk$-resolved spectral function $A(\vk,\omega)$  plotted in Fig.~\ref{fig:Akw_Fe_all_phases} for
the temperature of 5800~K. The bcc phase features 
a low-energy $e_g$ band along the $N-\Gamma-P$ path that is strongly broadened, thus 
indicating destruction of quasiparticle states. The nFL behavior of $e_g$ states in $\alpha$-Fe is explained by the narrow peak in its partial density of states (PDOS) induced by a van Hove singularity in the vicinity of $E_F$.
Such narrow peak in PDOS located at $E_F$ leads to suppression of the low-energy hopping and to the corresponding enhancement of correlations \cite{Mravlje2011}. 
In hcp Fe the electronic states in the vicinity of $E_F$ are sharp (their
red color indicating high value of $A(\vk,\omega)$), hence $\epsilon-$Fe exhibits a typical behaviour of a FL
with large quasi-particle life-times in the vicinity of $E_F$.  $\gamma$-Fe is in an intermediate state, with some broadening
noticeable in the $e_g$ bands at $E_F$ in the vicinity of the $\Gamma$ and $W$ points.

The conclusion of Ref.~\cite{Pourovskii2013} on the FL nature of $\epsilon$-Fe was subsequently challenged by Zhang {\it et al.}~\cite{Zhang2015}, who recalculated  $\epsilon$-Fe at the EIC volume within DFT+DMFT \footnote{Ref.~\cite{Zhang2015} was subsequently retracted by the authors \cite{Zhang2016} due to a numerical mistake in their transport calculations. However, this retraction does not concern their conclusions on a nFL nature of $\epsilon$-Fe at the EIC conditions.} and found a strongly nFL linear dependence of $\Gamma$ vs. $T$. In contrast to Ref.~\cite{Pourovskii2013} employing the density-density approximation to the local Coulomb vertex defined by $U=$3.4~eV and $J_H=$0.94~eV, Zhang {\it et al.} used the full rotationally-invariant form for the vertex parametrized by a higher value of $U=5$~eV and almost the same $J_H$. Therefore, in order to convincingly establish  the nature of electronic state in $\epsilon$-Fe Ref.~\cite{Pourovskii2017} performed new DFT+DMFT calculations for the $\epsilon$-phase with the full rotationally-invariant Coulomb interaction and explored the range  of $U$ from 4 to 6~eV. These calculations predicted an almost perfectly quadratic FL temperature scaling of $\Gamma$.

\begin{figure}[!t]
	\begin{center}
		\includegraphics[width=0.5\textwidth]{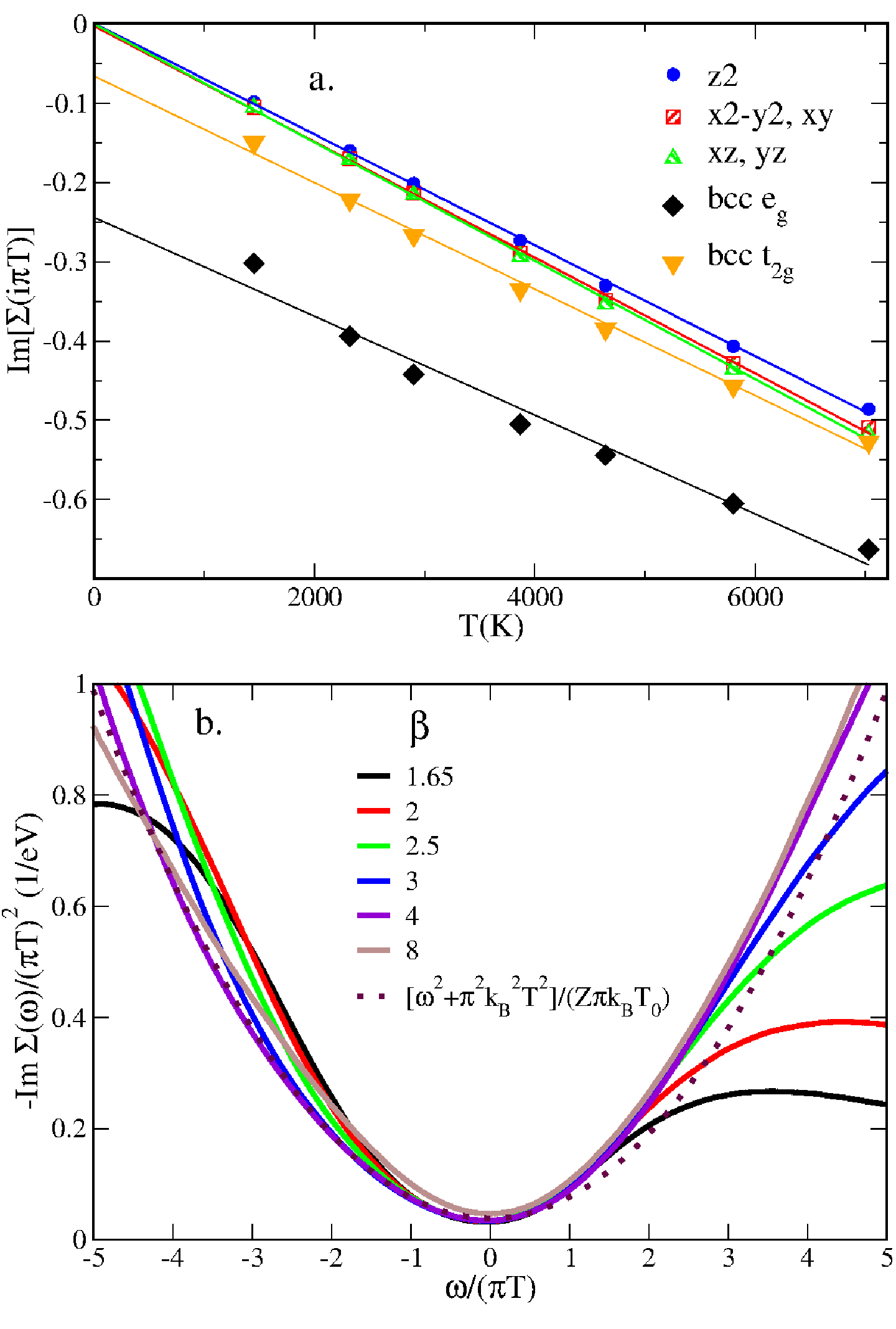}
	\end{center}
	\caption{Fermi-liquid scaling of the DMFT self-energy in $\epsilon$-Fe. a. The imaginary part of the DMFT self-energy at the first Matsubara point $\omega_1=i\pi k_B T$ vs. temperature for hcp and bcc
		Fe. Note that $Im[\Sigma(i\pi k_B T)]$ being proportional to T is a signature of a Fermi-liquid \cite{Chubukov2012} . The
		lines are the linear regression fits to the calculated points for corresponding 3$d$ orbitals of Fe.
		b. The rescaled imaginary part of the DMFT self-energy at the real axis $Im[\Sigma(\omega)]/(\pi k_B T)^2$ 
		vs. $\omega/(\pi k_B T)$. One sees that all self-energies collapse into a single curve described by a parabolic
		fit (the dotted line) defined by the quasiparticle weight $Z=$0.7 and the characteristic Fermi-liquid
		temperature scale $T_0=$12~eV. Adapted from Ref.~\cite{Pourovskii2017}.} 
	\label{fig:epsFe_FL_tests}
\end{figure}

A significant problem in the analysis of DFT+DMFT results carried out in Refs.~\cite{Pourovskii2013,Zhang2015} stems from the fact that the DMFT self-energy is calculated by CT-QMC on imaginary-frequency Matsubara points. The analytical continuation needed to obtain real-frequency data from this imaginary-frequency self-energy $\Sigma(i\omega)$ is known to be a  mathematically ill-defined problem and quite sensitive to the details of its implementation.  Even the extrapolation of $\Sigma(i\omega)$ to $\omega=$0 needed to evaluate $\Gamma$, eq.~\ref{eq:Gamma}, becomes rather less reliable for high temperatures, where the first Matsubara point $\omega_1=i\pi k_B T$ is  shifted away significantly from the real axis. 

Hence, Ref.~\cite{Pourovskii2017} also assessed  the FL nature of $\epsilon$-Fe by analyzing the imaginary-frequency self-energy without resorting to any analytical continuation. This is done by employing the so-called "first-Matsubara-frequency" rule. As demonstrated, e. g., by Chubukov and Maslov \cite{Chubukov2012}, in a Fermi liquid the imaginary
part of electronic self-energy at the first Matsubara point within a local approximation like
DMFT must be proportional to the temperature, i.e.  $Im[\Sigma(i\pi k_B T)]=\lambda T$, where $\lambda$ is a real
constant. In Fig.~\ref{fig:epsFe_FL_tests}a $Im[\Sigma(i\pi k_B T)]$ is plotted as a function of temperature for all
inequivalent orbitals in hcp and bcc Fe. One clearly sees that in the $\epsilon$ phase $Im[\Sigma(i\pi k_B T)]$ is almost perfectly proportional to $T$, in contrast to bcc Fe, where it exhibits significant
deviations from the "first-Matsubara-frequency" rule. This deviation is especially pronounced for the $e_g$ states of the bcc phase, which are indeed of a strongly nFL nature, as discussed above. 

Pourovskii {\it et al.}~\cite{Pourovskii2017} also verified the scaling of the full analytically-continued DMFT self-energy , which in a FL state exhibits the quadratic frequency dependence at low $\omega$ with $\Sigma(\omega) = C \cdot (\omega^2+(\pi k_B T)^2)$. The constant of proportionality $C$ can be written as $1/(Z \pi k_B T_0)$ with the characteristic scale $T_0 \sim 10T_{FL}$, where $T_{FL}$ is the temperature where resistivity ceases to 
follow a strict $T^2$ temperature dependence \cite{Berthod2013}. Indeed, one sees in Fig.~\ref{fig:epsFe_FL_tests}b that the real-frequency self-energies for different temperatures collapse into a single curve when plotted as $Im[\Sigma(\omega)]/(\pi k_B T)^2$ vs. $\omega/(\pi k_B T)$. The  value of $k_B T_0=$ 12~eV extracted from this plot corresponds to a $T_{FL} \approx$14000~K, which is significantly higher than the range of temperatures expected inside the EIC. This analysis of both the Matsubara and real-ferquency self energy of $\epsilon$-Fe has thus  confirmed its FL state. We will see in Sec.~\ref{subsec:epsFe_transport}  that this results has a direct bearing on the transport properties of $\epsilon$-Fe at the EIC conditions.

\begin{figure}[!htbp]
	\begin{center}
		\includegraphics[width=0.6\textwidth]{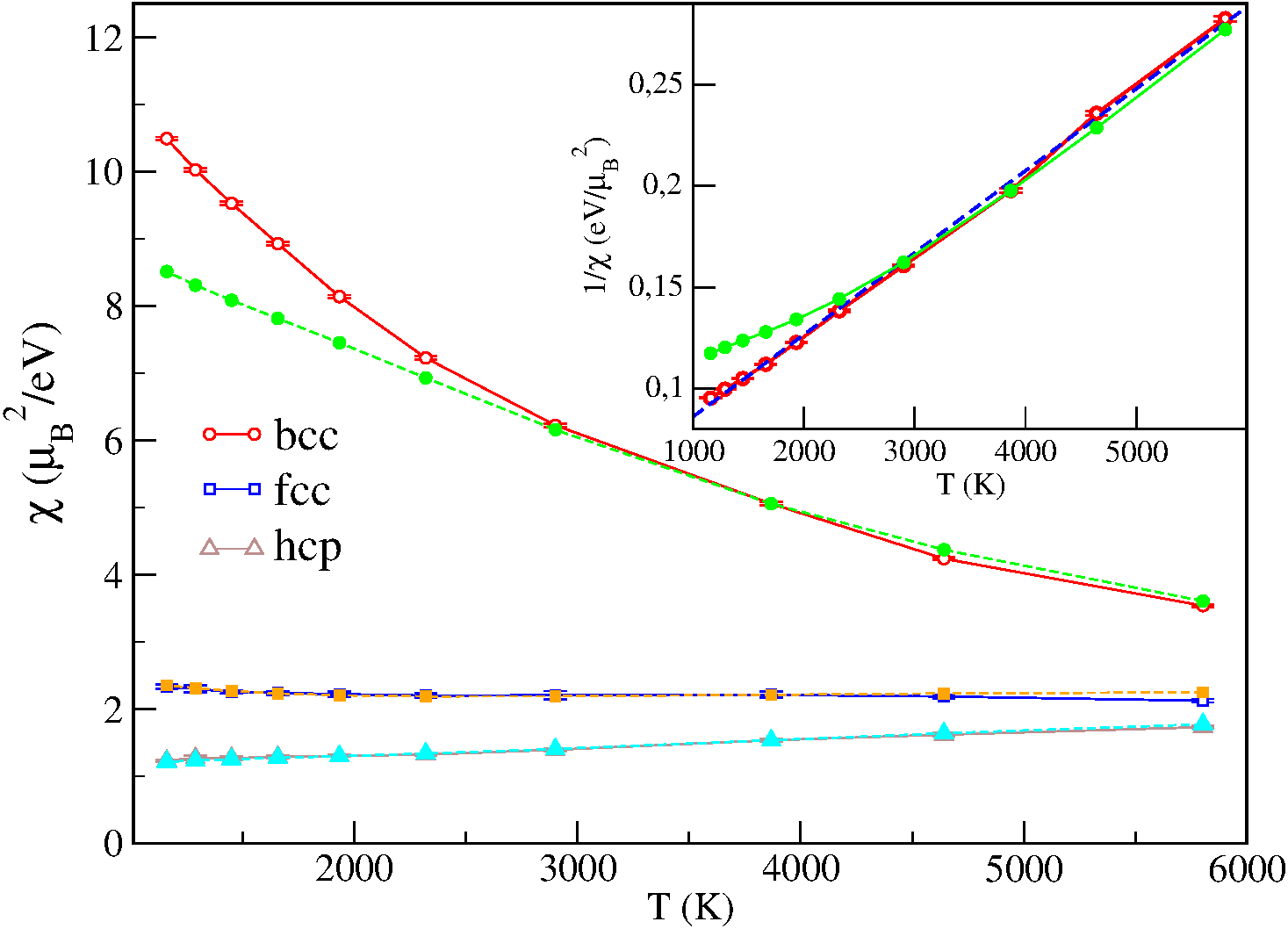}
	\end{center}
	\caption{The  uniform magnetic susceptibility in paramagnetic state
		versus temperature.  The error bars are due to the CT-QMC stochastic error. The  dashed lines with corresponding filled symbols are fits to the enhanced Pauli law, see the text. Inset: the inverse uniform magnetic suscptibility of bcc Fe is shown in red (empty circles), the blue dot-dashed and green (filled circles) lines are fits to the Curie-Weiss and enhanced Pauli law, respectively. Adapted from Ref.~\cite{Pourovskii2013}.} 
	\label{fig:Fe_uniform_susc}
\end{figure}

The temperature dependence of uniform susceptibility $\chi(T)$ was also calculated by Ref.~\cite{Pourovskii2013}  by evaluating the response to a small external field. The obtained temperature dependence (see  Fig.~\ref{fig:Fe_uniform_susc}) is consistent with the results on electronic structure discussed above. A Pauli behavior found for the FL $\epsilon$ and  $\gamma$ phases, while the nFL bcc $\alpha$ exhibits a Curie-Weiss behavior well described by the fit $\chi=\frac{1}{3}\frac{\mu_{eff}^2}{T+\Theta}$ with $\mu_{eff}=$2.6~$ \mu_B$ and $\Theta=$1396~$K$ (see inset in Fig.~\ref{fig:Fe_uniform_susc}). Alternatively, one may try to account for the same dependence with an enhanced Pauli law,  $\chi=\chi_0/(1-I * \chi_0)$, where $I$ is the Stoner parameter and $\chi_0$ is the bare susceptibility of Kohn-Sham band structure; the strong temperature dependence of  $\chi$ is then caused by a narrow peak at $E_F$ in the $e_g$ PDOS due to the van Hove singularity. However, the enhanced Pauli-law fit describes $\chi(T)$ less well than the Curie-Weiss one, the difference is clear for lower $T$ below 2500~K. Hence, from these calculations one may infer the existence of a rather large local magnetic moment in the bcc phase at the EIC conditions.
 One may expect a significant contribution to the $\alpha$-phase free energy due to the corresponding magnetic entropy.
 
Ruban {\it et al.}~\cite{Ruban2013} also studied the stability of local moments in iron at the EIC atomic volume using a longitudinal spin-fluctuation model employing first-principles intersite exchange interactions. They predicted a local moment of approximately the same magnitude to be stable in all three phases at the EIC temperature of 6000~K and also obtained an evolution of $\chi$ vs. $T$ that is qualitatively similar to the one of Ref.~\cite{Pourovskii2013}. They explained the qualitative difference in  $\chi(T)$ between the three phases by an impact of intersite pair interactions. Vekilova {\it et al.}~\cite{Vekilova2015} has subsequently studied the DMFT local susceptibility $\chi_{loc}$ , i.~e., the response to a local field applied to a single iron site, vs. $T$. They found a Curie-like temperature evolution in the bcc phase and a Pauli-like quasi temperature independent $\chi_{loc}$  in hcp-Fe. 
One may notice that within the single-site DMFT approximation $\chi_{loc}$ cannot be affected by intersite interactions. Hence, a qualitatively different behavior of $\chi_{loc}$  in the three phases hints at the key role of local correlation effects, in particular, existence of a local moment in bcc-Fe and its absence in the hcp phase.

\subsection{Impact of nickel substitution on electron correlations}\label{subsec:FeNi_in_core_elstruct}

The EIC is expected to contain, apart of iron, also non-negligible contributions of other transition metals, mainly of nickel as evidenced by the composition of metallic meteorites. The contribution of nickel is evaluated to 5-10\% based on geochemical models \cite{McDonough1995}. FeNi alloys have thus been intensively probed in laser-heated anvil cell experiments with some studies observing an extended $T-P$  region of  the fcc phase in Fe-rich FeNi alloys, with the $\epsilon$ phase still projected, however, to be stable at the EIC conditions \cite{Lin2002_1,Kuwayama2008,Mao2006,Sakai2011,Tateno2012}. In contrast,   a bcc phase was observed  in Fe$_{92}$Ni$_8$ at high $T$ and $P$ by Dubrovinsky {\it et al.}~\cite{Dubrovinsky2007}.  The relative stability of fcc, bcc, and hcp phases for Fe-rich FeNi alloys at the EIC conditions has been considered in several theoretical works \cite{Ekholm2011,Cote2012,Martorell2013} using DFT-based methods and thus neglecting dynamical many-electron effects. As shown in the previous section, these effects are qualitatively different in the three phases in the case of pure Fe.

The impact of Ni substitution on many-electron effects in these phases  at the EIC conditions is, hence, an important subject and has been studied  by means of DFT+DMFT~\cite{Vekilova2015, Hausoel2017}. In particular, Vekilova {\it et al.}~\cite{Vekilova2015} employed the same computational framework as Ref.~\cite{Pourovskii2013} and modeled the random Fe$_3$Ni alloy by the smallest supercells capable to accommodate 25\% of Ni substitution. These supercells comprise two, one, and two conventional cells in the case of bcc, fcc, and hcp lattices, respectively. In order to model  more realistic lower Ni concentrations one would have to employ larger supercells with the corresponding heavy increase in the computational effort.  In addition, Vekilova {\it et al.} made use of different environment of two inequivalent Fe sites in their bcc and hcp supercells , with  only one of those having Ni nearest neighbors, to evaluate the effect of  Ni nearest neighbors on correlations on iron sites. Many-electron effects on Ni were included in the same way as for Fe in Ref.~\cite{Pourovskii2013} with the corresponding local Coulomb interaction specified by the same values of $U=3.4$~eV and $J_H=$0.9~eV. 

\begin{figure}[!t]
	\begin{center}
		\includegraphics[width=0.8\textwidth]{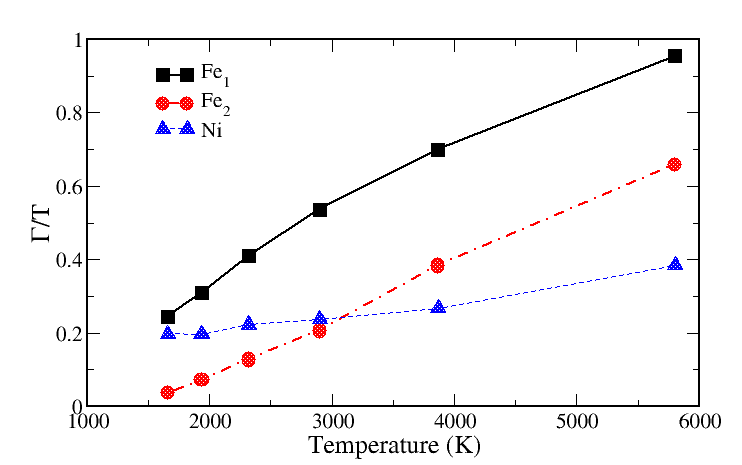}
	\end{center}
	\caption{The inverse quasiparticle lifetime $\Gamma$ as a function of $T$ for three inequivalent sites, Fe$_1$ (six Ni and six Fe nearest neighbors) , Fe$_2$ (all nearest neighbors are Fe) and Ni, in the hcp Fe$_3$Ni supercell.} 
	\label{fig:Fe3Ni_hcp_gammas}
\end{figure}

The effect of Ni nearest neighbors (NN) on electronic correlations on Fe sites was found to be structure-dependent. In the bcc phase it results in significant deviations from the Curie-Weiss behavior for the uniform susceptibility $\chi$ and reduced $\Gamma$ for the $e_g$ states. Overall the presence of Ni NNs reduced the degree of "non-Fermi-liquidness" for the bcc $e_g$ states. The opposite effect was found for the hcp phase, where  the presence of Ni NNs enhanced the uniform susceptibility and $\Gamma$ (Fig.~\ref{fig:Fe3Ni_hcp_gammas}).  These effects can be related to modifications of corresponding Fe PDOS due to the presence of Ni NNs. Namely, in the case of bcc  one observes a smearing of the $e_g$ peak at $E_F$, conversely, in the case of $\epsilon$-Fe a characteristic dip in PDOS in the vicinity of $E_F$ becomes more shallow. 

Vekilova {\it et al.} found rather weak correlation effects on Ni sites  at the EIC conditions. As shown in Fig.~\ref{fig:Fe3Ni_hcp_gammas}, $\Gamma$ for Ni features a nFL behavior with a rather slow increase in the studied range of $T$. 

Many-electron effects in Ni  and FeNi alloys under extreme conditions were subsequently studied in a recent work by Hausoel {\it et al.}~\cite{Hausoel2017}. The authors employed a DFT+DMFT technique that is similar to the one used in Refs.~\cite{Pourovskii2013,Vekilova2015} and mainly focused on nFL properties of  Ni  3$d$ states, this question was not addressed by the previous works. They modeled random Fe$_{1-x}$Ni$_x$ alloys ($x=$0.05, 0.20) at the EIC density within the coherent-potential approximation (CPA). The advantage of CPA is that one can treat any concentration $x$ with the same computational cost, however,   the local environment effects, which seems to be quite important, as one sees in Fig.~\ref{fig:Fe3Ni_hcp_gammas}, are neglected.  Hausoel {\it et al.}  predicted a strong enhancement of $\Gamma$  due to the Ni substitution as compared to pure $\epsilon$-Fe for the studied range of temperatures up to 2000~K. This result is in agreement with Fig.~\ref{fig:Fe3Ni_hcp_gammas}, if one compares the magnitude of $\Gamma$ for the iron site Fe$_2$ without Ni NNs with that for Ni at $T <$2000~K. However, one also sees that $\Gamma$ of Ni exhibits a slow almost linear-in-T scaling, while $\Gamma$ of Fe$_2$ scales quadratically with $T$, hence at the EIC temperature of about 6000~K the scattering due to the iron sites dominates and the Ni  contribution is relatively weaker.

\subsection{Electron-electron scattering and transport in $\epsilon$-Fe}\label{subsec:epsFe_transport}

Transport properties of iron at the extreme conditions are of significant importance for geophysics. In particular, the thermal conductivity of the iron-rich matter inside the liquid outer core of Earth is a key parameter determining the stability of the geodynamo generating the Earth's magnetic field. This geodynamo runs on heat from the growing solid inner core and on chemical convection provided by light elements issued from the liquid outer core on solidification~\cite{Pozzo2012}. The power supplied to drive the geodynamo is proportional to the rate of inner core growth, which in turn is controlled by heat flow at the core-mantle
boundary~\cite{Lay2008}.  This heat flow critically depends on the thermal conductivity of liquid iron under the extreme pressure and temperature conditions in the Earth's core.  For a long
time there has been agreement that convection in the liquid outer core
provides most of the energy for the geodynamo since at least 3.4 billion years~\cite{Olson2013,Stacey2007}. 
Recently, such a view has been challenged by first-principles
calculations~\cite{deKoker2012,Pozzo2012}, suggesting a much higher
capacity for the liquid core to transport heat by conduction and
therefore less ability to transport heat by
convection~\cite{Olson2013}.  The calculated conductivities have been
found to be two to three times higher than the  earlier generally accepted
estimates.

Convection also plays a crucial role in the current theory of the
EIC dynamics, as a radial motion of the inner core matter is
invoked to explain the observed seismic anisotropies of the inner
core~\cite{Romanowicz1996,Buffett2009,Monnereau2010}. However, {\it ab
	initio} calculations of Ref.~\cite{Pozzo2014} similarly predict a too high
thermal conductivity for hexagonal close-packed (hcp) $\epsilon$-iron to sustain this convection. Experimental measurements of the thermal conductivity at the core conditions has not been reported to date, though two anvil-cell results above $P=$100 GPa and at $T$ of several thousands K have been recently reported.  Ohta {\it et al.}~\cite{Ohta2016} measured the electrical resistivity and then obtained the thermal conductivity using the Wiedemann-Franz law with  the standard Lorenz number; in contrast,  Kon\^{o}pkov\'a {\it et al.} ~\cite{Konopkova2016} directly extracted the thermal conductivity from propagation of heat generated on one side of the sample by a laser pulse. The resulting  thermal conductivities, extrapolated to the core-mantle boundary conditions, differ  by almost an order of magnitude, with shock-wave measurements extrapolation of Refs.~\cite{Matassov1977,Stacey2001,Stacey2007,Seagle2013} agreeing better with the value of $\sim$30 Wm$^{-1}$K$^{-1}$ reported by Kon\^{o}pkov\'a {\it et al.} In contrast, the value above 200 Wm$^{-1}$K$^{-1}$ reported Ohta {\it et al.} lends support to the DFT predictions of  Refs.~\cite{deKoker2012,Pozzo2012,Pozzo2014}.  Overall, the magnitude of thermal conductivity in both liquid and solid iron at the core's conditions remain an unresolved problem  in geophysics, see Ref.~\cite{Williams2018} for a recent review.

These first-principles calculations for
liquid and solid iron \cite{deKoker2012,Pozzo2012,Pozzo2014}
employed the standard density-functional-theory (DFT)
framework in which electron-electron repulsion is not properly
accounted for as dynamical many-body effects are neglected. 
Hence, the contribution to resistivity from the electron-electron scattering (EES) of $d$-electrons due to correlations was not taken into account in those calculations. In order to elucidate how large the EES contribution to the electrical
and thermal resistivity at Earth's core conditions Pourovskii {\it et al.}~\cite{Pourovskii2017} extended their  DFT+DMFT approach to calculations of the electrical and thermal conductivities of pure $\epsilon$-Fe at the EIC density. Using the analytically-continued DMFT self-energy (see Fig.~\ref{fig:epsFe_FL_tests}b)  they evaluated  the conductivity from the corresponding DFT+DMFT spectral function using the Kubo linear-response formalism \cite{Kotliar2006,triqs_dft_tools} and neglecting vertex corrections. Namely, the electrical and thermal conductivity read

\begin{equation}
	\sigma_{\alpha\alpha'}=\frac{e^2}{k_BT}K^0_{\alpha\alpha'}, 
\end{equation}
\begin{equation}
	\kappa_{\alpha\alpha'}=k_B\left[K^2_{\alpha\alpha'}-\frac{(K^1_{\alpha\alpha'})^2}{K^0_{\alpha\alpha'}}\right],
\end{equation}
where $\alpha$ is the direction ($x$, $y$, or $z$), $k_B$ is the Boltzmann constant.  The kinetic coefficients $K^n_{\alpha\alpha'}$ can be calculated from the real-energy DFT+DMFT spectral function $A(\vk,\omega)$ and the velocities of Kohn-Sham states, $v_{\alpha}(\vk)$,  the later is evaluated by DFT band structure methods as described, e.~g., Ambrosch-Draxl {\it et  al.}~\cite{Ambrosch-Draxl2006} for the case of LAPW method.

\begin{figure}
	\begin{center}
		\includegraphics[width=0.8\textwidth]{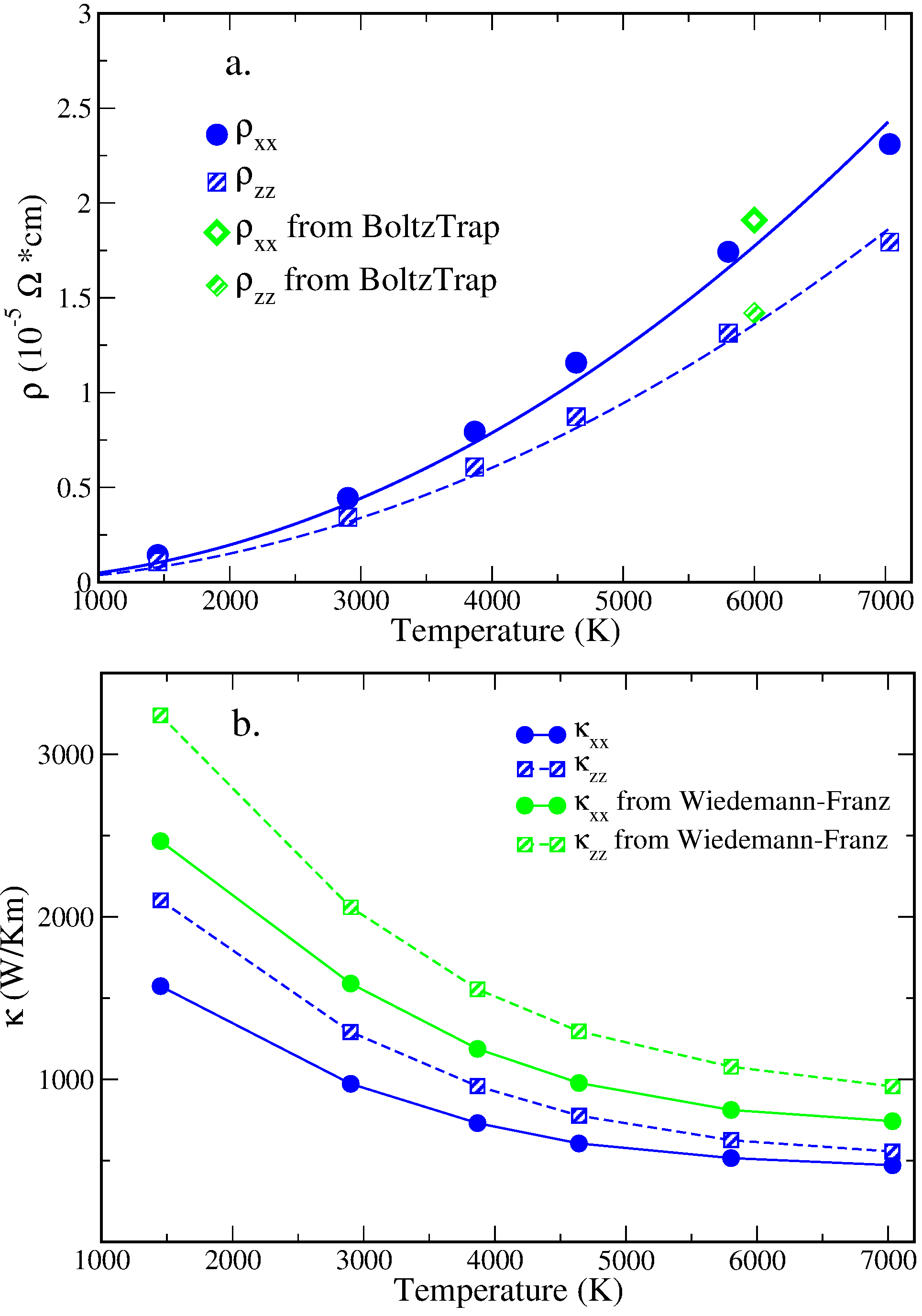}
	\end{center}
	\caption{\label{fig:conduct} Calculated electron-electron-scattering contribution to the electrical and thermal
		resistivity of hcp iron at Earth's core density. a. Electrical resistivity. Blue filled circles and hashed squares are DFT+DMFT results for $\rho_{xx}$ and $\rho_{zz}$ , respectively. Green empty and hashed diamonds are the corresponding resistivities calculated by the Boltzmann-transport code BoltzTrap \cite{Madsen2006} assuming a
		Fermi-liquid with the scattering rate $\Gamma/Z=$0.09~eV. 
		b. Thermal conductivity. Blue
		filled circles and hashed squares are DFT+DMFT results for $\kappa_{xx}$ and $\kappa_{zz}$, respectively. Green lines/symbols are the corresponding conductivities obtained from the calculated
		electrical conductivity using the Wiedemann-Franz law with the standard Lorenz number of 2.44$\cdot$10$^{-8}$~W$\Omega$K$^{-2}$. Adapted from Ref.~\cite{Pourovskii2017}.
	}
\end{figure}

The contributions of electron-electron scattering into the electrical resistivity and thermal conductivity of $\epsilon$-Fe obtained by  Ref.~\cite{Pourovskii2017} are displayed as a function of $T$ in Figs.~\ref{fig:conduct}a and \ref{fig:conduct}b, respectively. First, one sees that the electrical resistivity $\rho$ features a clear $T^2$ FL dependence, as expected on the basis of the analysis of its DMFT self-energy as discussed in Sec.~\ref{subsec:Fe_in_core_elstruct}. Second, its magnitude of  1.6$\cdot$ 10$^{-5}$~$\Omega \cdot$cm at $T=$5800~K is rather insignificant compared to the electron-phonon-scattering contribution of about 5.3$\cdot$10$^{-5}$~$\Omega \cdot$cm predicted by DFT calculations of Pozzo {\it et al.}~\cite{Pozzo2014}. This  indicates that the electron-electron
scattering should not strongly influence the electrical resistivity in hcp-Fe at EIC conditions. Third, the  electron-electron-scattering thermal conductivity $\kappa_{e-e}$  of  540~Wm$^{-1}$K$^{-1}$ $T=$5800~K is not high and comparable to the corresponding value due to the electron-phonon scattering $\kappa_{e-ph}\approx$300~Wm$^{-1}$K$^{-1}$ obtained by Ref.~\cite{Pozzo2014}. Hence, in contrast to  $\rho$  the electron-electron scattering contribution to the thermal conductivity is quite important. By including both the electron-electron and electron-phonon scattering effects the total conductivity is reduced to about 190~Wm$^{-1}$K$^{-1}$, hence, the corresponding resistivity is enhanced by about 60\%. 

An important observation of Ref.~\cite{Pourovskii2017}  is that the DFT+DMFT electron-electron-scattering thermal conductivity of $\epsilon$-Fe is significantly lower than the one calculated from the corresponding contribution to $\rho=1/\sigma$ in accordance with the Wiedemann-Franz law, $\kappa/(\sigma T)=\frac{\pi^2}{3}\left(\frac{k_B}{e}\right)^2=L_0$ (where the standard Lorenz number  $L_0$ is 2.44$\cdot 10^{-8}$W$\Omega$K$^{-2}$), see Fig.~ \ref{fig:conduct}b. By  employing simple analytical calculations in the Boltzmann formalism Herring~\cite{Herring1967} showed that the quadratic FL frequency dependence of the imaginary part of the self-energy and, hence, of the quasiparticle life-time 
$$
1/\tau(\epsilon)=1/\tau(\epsilon=0)\cdot \left(1+\epsilon^2/(\pi k_B T)^2\right),
$$
leads to a substantial reduction of the Lorenz number 
$$
\kappa/(\sigma T)=L_0/1.54=L_{FL}.
$$
The stronger effect of the frequency-dependence of $\tau(\omega)$ on the thermal conductivity as compared to $\sigma$ 
is due to the additional power $\epsilon^2$ in the numerator of the transport integrals for $\kappa$, see, for example, Ref.~\cite{AshcroftMermin}. Hence, the enhancement of the electron-electron-scattering contribution to the thermal resistivity obtained within DFT+DMFT stems directly from the Fermi-liquid state of the $\epsilon$-Fe phase at the EIC conditions. A similar reduction of the Lorenz number has been very recently obtained by another DFT+DMFT study \cite{Xu2018}; the electron-electron contribution to the thermal conductivity of $\epsilon$-Fe at the EIC conditions reported in this works is close to that of Ref.~\cite{Pourovskii2017}.

The reduction of the thermal conductivity due to the electron-electron scattering predicted by Ref.~\cite{Pourovskii2017} is still insufficient to explain the stability of convection by itself. On the other hand, the extremely low values of $\kappa_{tot}\sim$50 Wm$^{-1}$K$^{-1}$ may not be required to reconcile theoretical calculations of the thermal conductivity with geophysical observations \cite{ORourke2016,Hirose2017}. 

Moreover, the impact of alloying and  lattice vibrations have not been to date  taken into account in the DFT+DMFT transport calculations.  For example, the DFT+DMFT calculations for Fe-Ni alloy at the inner core conditions discussed in the previous section  point out an important local environment effects that may affect the electron-electron scattering in real material of the EIC. The  impact of all those effects on transport properties of the EIC matter remains to be evaluated.

\subsection{Many-electron effects and structural stability}\label{subsec:Fe_free_energy}

The stable phase of pure iron at the EIC conditions has not been clearly identified experimentally; neither have {\it ab initio} DFT calculations resulted in an unambiguous prediction due to a small energy difference between the three phases, as described in the beginning of Sec.~\ref{sec:Fe_in_core}. Hence, corrections due to the many-electron effects neglected by DFT can have a qualitative impact on the nature of stable iron phase at the EIC conditions.

A quantitative estimation for the contribution of correlations to the electronic free energy  of the three phases   was obtained  by Ref.~\cite{Pourovskii2013} together with their other magnetic and electronic properties (see Sec.~\ref{subsec:Fe_in_core_elstruct}). Their fixed-lattice calculations neglected the contribution of lattice vibrations to the phase stability, which are expected to be very significant at such extreme temperatures. However, such calculations are still able to assess the structural dependence of this contribution. 

In spite of the simplifying fixed-lattice approximation evaluating the electronic free energy within the DFT+DMFT framework remains a highly non-trivial task. The total-energy calculations in this framework have nowadays become quite standard as described in Appendix.~\ref{sec:etot}. Such DFT+DMFT calculations evaluating the total energy using eq.~\ref{eq:E_DMFT} have been applied, for example,  by Leonov~{\it el al.}~\cite{Leonov2011} to study the $\alpha$-$\gamma$ phase transition in iron.

\begin{figure}[!t]
	\begin{center}
		\includegraphics[width=0.65\columnwidth]{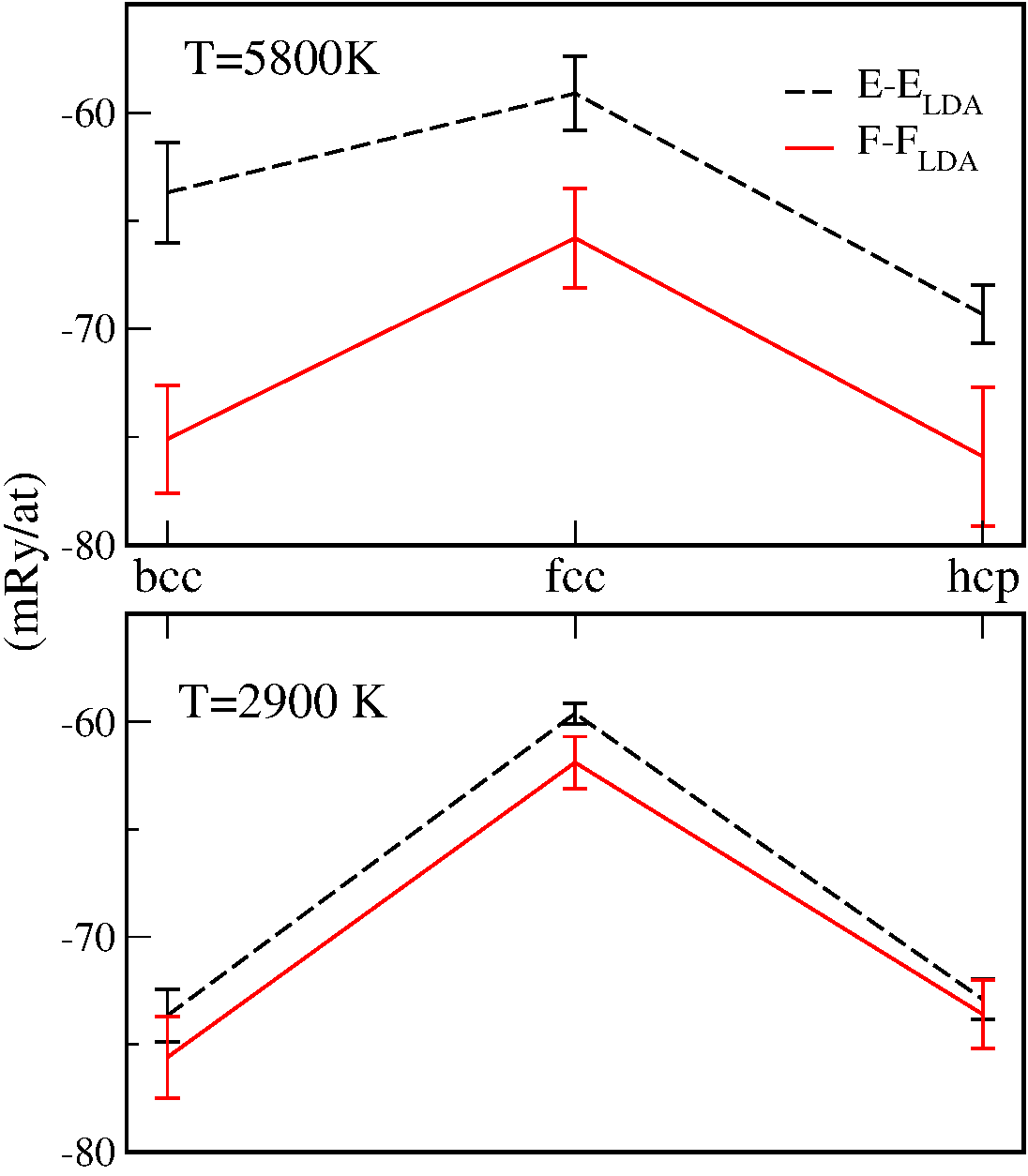}
		\caption{\label{fig:free_enr}
			Many-body correction to the total (black dashed line) and free (red solid line) energy for the
			three phases of Fe at the volume of 7.05 \AA/atom at T=5800~K (upper panel) and 2900~K (lower panel).
			The error bars are due to the CT-QMC stochastic error. Adapted from Ref.~\cite{Pourovskii2013}.
		}
	\end{center}
\end{figure}

In contrast, the partition function and, correspondingly, the DFT+DMFT grand potential (\ref{dmft_gp}) cannot be generally directly sampled by the usual Metropolis algorithm. In the context of DMFT quantum impurity problem solved by CT-QMC or other numerical technique, it is the contribution of DMFT functional  $\Phi_{imp}[G_{loc}({\bf R})]$  into (\ref{dmft_gp}), which is the sum of all local skeleton diagrams constructed with the local GF $G_{loc}(i\omega_n)$ and the on-site vertex, that cannot be computed directly. Different types of the numerical thermodynamic integration are employed instead, in particular, the one from an analytical high-temperature limit \cite{Haule2015}. Such integration remains non-trivial in the present case of Fe at the EIC conditions, as the temperature $T\approx$6000~K is still low compared to other energy scales like the bandwidth or $U$. Ref.~\cite{Pourovskii2013} employed instead the numerical thermodynamic integration over the coupling strength $\lambda \in [0:1]$, where the corresponding free energy is defined $F_{\lambda}=-\frac{1}{\beta}\ln \Tr\left(\exp[-\beta(\hat{H}_0+\lambda \hat{H}_{int})]\right)$, $H_0$ is the one-electron  part of the DFT+U Hamiltionian (\ref{eq:H_DFTU}), $\hat{H}_{int}=\hat{H}_U-E_{DC}$ is the interacting part. The coupling constant integration results in the following expression for the many-body correction:

\begin{equation}\label{eq:dF}
	\Delta F= F-F_{DFT} = \int_0^1 \frac{\langle \lambda \hat{H}_{int} \rangle_{\lambda}}{\lambda} d\lambda,
\end{equation} 
where $F_{DFT}$ is the electronic free energy in DFT. In derivation of Eq.~\ref{eq:dF} one neglects the $\lambda$  dependence of the one-electron part, and, hence, the charge density renormalization due to many-body effects. In practice, the integrand in (\ref{eq:dF}) was evaluated numerically with $\frac{\langle \lambda  \hat{H}_{int}  \rangle_{\lambda}}{\lambda}$ computed for a discrete mesh in $\lambda \in [0:1]$ by performing DFT+DMFT simulations with the Coulomb interaction scaled accordingly.  This method was subsequently applied in DFT+DMFT calculations of Bieder {\it et al.}~\cite{Bieder2014} to evaluate the free energy of the cerium metal.

The resulting DMFT correction to the free energy for the three iron phases is plotted in Fig.~\ref{fig:free_enr} togehter with the correction to the total energy calculated given by the difference of (\ref{eq:E_DMFT}) and $E_{DFT}$. Within rather significant error bars the magnitude of  $\Delta F$ is the same for bcc and hcp Fe, which are suggested as stable phases of iron~\cite{Vocaldo2003} and iron-based alloys \cite{Lin2002,Dubrovinsky2007} at the Earth's inner core conditions.  Though the correlation strength (as measured, for example, by the inverse quasiparticle lifetime $\Gamma$,   Fig.~\ref{fig:Gamma_Fe_all_phases}), is higher for $\alpha$-Fe, this is apparently compensated by a higher value of $U$ predicted for the $\epsilon$-phase by cRPA calculations of the same work~\cite{Pourovskii2013}. The magnitude of $\Delta F$ is, however, at least several mRy smaller in the case of fcc Fe, showing that the many-body correction may significantly affect relative energy differences among iron phases at the Earth core conditions.
One may also notice that the entropic contribution $T\Delta S =  \Delta E - \Delta F$ becomes much more significant at the higher temperature, and its contribution is almost twice larger in the case of the bcc phase compared with two others. This is in agreement with the local-moment behavior of this phase predicted by DFT+DMFT calculations, as detailed above. 

\section{Conclusions and perspectives }\label{sec:Fe_conclusions}

In this Topical review we have discussed  many-electron effects in various phases of iron, with particular emphasis on the results obtained by recent {\it ab initio} DFT+DMFT calculations. We concentrated on two particularly interesting regions of the Fe $P-T$ phase diagram: (i) the region of moderate pressure up to 60 GPa, around the $\alpha\to\epsilon$ transition, where the hexagonal $\epsilon$ phase exhibits exotic and poorly understood properties; and (ii)  the geophysically relevant region of pressure about 360 GPa and temperature of about 6000~K.   

In the moderate-pressure region (i) we predict a significant enhancement of dynamical correlations at the pressure-induced $\alpha \to \epsilon$ phase transition. This enhancement is explained by the fact that dynamical correlations are strongly suppressed by the static spin polarization in ferromagnetic $\alpha$-Fe; this polarization is absent in paramagnetic hcp $\epsilon$-Fe. In result, DFT+DMFT calculations predict large many-body corrections to the equation of state of the $\epsilon$-phase and a significant electron-electron scattering contribution to its electrical resistivity. The same theoretical framework predicts an electronic topological transition to occur in this intermediate pressure range thus explaining the observed peculiarities in the evolution of its hexagonal cell parameters, Debye velocity and M\"ossbauer central shift. 

Applying the same framework to the geophysically-important regime (ii)  of the Earth's inner core (EIC) conditions  one finds a strong structural dependence for many-electron effects. Namely, bcc $\alpha$-Fe exhibits a clearly non-Fermi liquid behavior as evidenced by a sub-linear temperature dependence of the quasiparticle scattering rate $\Gamma$ and a Curie-Weiss-like behavior of the magnetic susceptibility. In contrast, an almost perfectly Fermi-liquid state is predicted for $\epsilon$-Fe at the same EIC conditions, with sharp quasiparticle bands at low-energy and the $T^2$ scaling of $\Gamma$. The fcc $\gamma$ phase is found to be in an intermediate regime between bcc and hcp with some noticeable deviations from Fermi-liquid state. The contribution of correlation effects to the electronic total and free energies is consequently also strongly structurally-dependent. The strength of many-electron effects on iron is found to be sensitive to the local environment and quite significantly affected by the presence of Ni nearest neighbors; our calculations also show comparatively weaker correlations on Ni sites themselves at the EIC conditions. Finally and quite unexpectedly, the  predicted "dull" Fermi-liquid state of $\epsilon$-Fe leads to a significant suppression of the Lorenz number with the corresponding enhanced contribution of the electron-electron scattering to the thermal resistivity. This enhancement of the thermal resistivity as compared to electrical one is directly related to a strong (quadratic) frequency dependence of the Fermi-liquid electron-electron scattering rate.

All these results have been obtained by the DFT+DMFT  in conjunction with the numerically-exact continuous-time quantum Monte Carlo method, which is equally reliable for all considered regimes of correlations (e. g., Fermi-liquid/non-Fermi-liquid, paramagnetic/ferromagnetic states). However, this technique, as well as overall DFT+DMFT framework, is numerically heavy and some approximations had to be employed to make these calculations feasible:

\begin{itemize}
	\item  Ref.~\cite{Pourovskii2014} employed the single-site DMFT in conjunction with a density-density local vertex. Though these calculations successfully accounted for the ground-state properties of the $\epsilon$ phase, the electron-electron scattering contribution into the resistivity was apparently strongly underestimated. The non-density-density terms of the local vertex were also found in this work to be essential to account for the collapse of static antiferromagnetism in this phase. The effect of rotationally-invariant local Coulomb repulsion in $\epsilon$-Fe thus needs to be fully investigated. A very significant contribution of the electron-electron scattering to the electrical resistivity of  $\epsilon$-Fe and its non-Fermi-liquid behavior at low temperatures, as well as the non-conventional (spin-fluctuation-pairing) superconductivity experimentally observed in this phase, also hint at important inter-site correlations, which can be included only by approaches beyond the single-site approximation. 
	
	\item The density-density approximation for the local vertex is probably less severe in the case of EIC conditions. In particular, the inclusion of rotationally-invariant local interaction in  $\epsilon$-Fe~\cite{Pourovskii2017} led only to some quantitative changes compared to the previous study within the density-density approximation (see also Appendix~\ref{chap:appendix_FeDensDens}). The non-local correlations are also expected to be less important away from magnetic instabilities and with a lower  strength of correlations at the  high-density of the EIC matter. In contrast, the fixed-lattice approximations is quite severe when one considers temperatures just below the melting.  Correspondingly, future studies of the impact of lattice vibrations on electronic correlations and vice versa are in this case of high importance. 
\end{itemize}

Hopefully, the recent progress in development of extended-DMFT frameworks (see  Rohringer {\it et al.}~\cite{Rohringer2018} for a recent review),  will eventually make accessible the most important  two-particle quantities (e.g., the full $\vk$ and $\omega$-dependent susceptibility or vertex corrections to the transport) for realistic multi-band systems with possibly significant intersite correlation, like $\epsilon$-Fe in the moderate pressure range. 

Regarding the second point: though fully consistent DFT+DMFT {\it ab initio} molecular dynamics will remain prohibitively computationally expensive for some time, one may still make use of the usual approximation of evaluating the electronic structure at fixed ionic coordinates. Hence, in order to assess the effect of lattice distortions on many-electron effects  one may employ a set of supercells representing characteristic deviations from the perfect atomic positions expected at the relevant temperature for a given phase.  Conversely, the impact of electronic correlations on lattice vibrations, at least in the harmonic approximation, can be studied using the recently formulated DFT+DMFT schemes for calculation of forces and phonon dispersions \cite{Leonov2012,Leonov2014_1,Haule2016}. Eventually, the impact of light elements inclusions needs to be also included in realistic simulations of the EIC matter.

\section*{Acknowledgments}
This review is mainly based on several recent works \cite{Pourovskii2013,Glazyrin2013,Pourovskii2014,Vekilova2015,Pourovskii2017} carried out by the author with a number of coauthors.  I want to especially acknowledge the valuable contribution of my colleagues theoreticians I. A. Abrikosov, A. Georges, J. Mravlje,  and S. I. Simak. I am also grateful to other coauthors of these papers:  M. Aichhorn, M. Ekholm, M. I. Katsnelson, T. Miyake, A. V. Ruban,  F. Tasn\'adi, O. Vekilova, V. Vildosola as well as our experimental collaborators from the group of L. Dubrovinsky. The many-body approaches employed in these papers were provided by the software library "TRIQS" that has been developed over years by  O. Parcollet,  M. Ferrero and collaborators. I thank J. Mravlje, S. Biermann,  M. Gabay, D. Givord, J. Kunes, M. Rozenberg, G. Sangiovanni, M. van Schilfgaarde for critical  reading of this manuscript and useful comments. The support of ANR and DFG under the collaborative project "RE-MAP" as well as of the European Research Council Grant No. ERC-319286-QMAC is gratefully acknowledged.

\appendix
\newpage
\begin{center}
	\Large{\bf Appendix}
\end{center}
\section{{\it Ab initio} dynamical mean-field theory approach: an overview}\label{dmft:overview}

The standard DFT framework is well known to be deficient in the case of partially-filled narrow bands; the effect of a  local Coulomb repulsion on the physics of such states cannot be captured by local or semi-local exchange-correlation (XC) functionals like the local-density or generalized-gradient approximations. As discussed in the introduction section, the 3$d$ band of iron, while having a bandwidth larger than the relevant local Coulomb interaction, still cannot be satisfactory captured within pure DFT. 

The approach employed for ab initio studies discussed in this review is thus based on supplementing the quadratic Kohn-Sham (KS) Hamiltonian $H_0$ with an explicit local Coulomb interaction between Fe 3$d$ states, the resulting "DFT+U" Hamiltonian \cite{Anisimov1991,Anisimov1997} reads

\begin{equation}\label{eq:H_DFTU}
\hat{H}_{DFT+U}=\hat{H}_0+\hat{H}_U-E_{DC}=\sum_{\vk \nu}\epsilon_{\vk\nu}c^{\dagger}_{\vk\nu}c_{\vk\nu}+\sum_{i,\atop 1,2,3,4}\langle 1 2|U| 3 4\rangle d^{\dagger}_{i 1} d^{\dagger}_{i 2} d_{i 4} d_{i 3}-E_{DC},
\end{equation}
where $c^{\dagger}_{\vk\nu}$($c_{\vk\nu}$) is the creation(annihilation) operator for the Kohn-Sham state $\psi_{\vk\nu}$  at $k$-point $\vk$ and the band index $\nu$, $d^{\dagger}_{i\alpha}$($d_{i\alpha}$) is the operator creating (annihilating)  localized states $w_{i\alpha}$ on the correlated  (3$d$) shell\footnote{For simplicity here and below we consider the case of a single correlated site per unit cell} in the unit cell $i$, $\alpha\equiv1,2,...$  is a compound index for relevant quantum numbers labeling one-electron orbitals within that shell (for example, $\alpha\equiv \{m\sigma\}$,  where $m$ is the orbital quantum number and $\sigma$ is the spin). The last term, $E_{DC}$, is the double-counting correction, which will be discussed below.

The interacting term in the DFT+U Hamiltonian is naturally defined in the real space, as the interaction is assumed to act between orbitals localized on the same atomic site. A sufficient localization of the orbitals $w_{i\alpha}$ at the correlated site is thus necessary for the DFT+U Hamiltonain to be physically sensible. For extended orbitals the intersite interactions are comparable to $U$; neglecting them in ($\ref{eq:H_DFTU}$) thus becomes a poor approximation \cite{Ayral2013,Hansmann2013}. However, in solids one cannot define $d$ or $f$ orbital as in  an isolated atom, as such definition makes sense near the nucleus, where the crystalline potential is approximately spherical, but not in the interstitial. 

There exists a number of approaches for constructing such bases representing  localized correlated states in solids.  
For example, one may employ the framework of Refs.~\cite{Marzari1997,Marzari2012,Anisimov2005,Lechermann2006,Amadon2008} and define the localized orbitals $w_{i\alpha}$ as Wannier functions constructed from a subset $\cal{W}$ of Kohn-Sham bands:
\begin{equation}\label{eq:WF}
w_{i\alpha}(\vr)=\sum_{\vk \in BZ} w_{\vk\alpha}(\vr+\vR_i)e^{-i\vk\vR_i}= \sum_{\vk \in BZ\atop v\in\cal{W} }e^{-i\vk\vR_i}\psi_{\vk\nu}(\vr+\vR_i)P_{\nu \alpha}(\vk),
\end{equation}
where the subset $\cal{W}$ comprises KS bands with a substantial contribution due to correlated orbitals, $\vR_i$ is the lattice vector of the unit cell $i$, $\hat{P}(\vk)$ is a complex matrix such that the resulting orbitals form an orthonormalized basis, $\langle  w_{i\alpha}|  w_{j\beta}\rangle=\delta_{ij}\delta_{\alpha\beta}$. In fact, matrices $\hat{P}(\vk)$ possessing such properties are well-known to be not uniquely defined, the resulting gauge freedom in $\hat{P}(\vk)$ can be exploited to obtain a well-localized basis of Wannier functions. Direct minimization of the spread of $w_{i\alpha}$ in the real space is employed  to construct the maximally-localized Wanniers basis \cite{Marzari1997}. Another, a projective construction of localized Wannier functions, avoiding the explicit spread minimization, was proposed by Amadon {\it et al.}~\cite{Amadon2008} and implemented in conjunction with the  linearized augmented planewave (LAPW) band structure method by Aichhorn {\it et al.}~\cite{Aichhorn2009}.
One may also mention a hybrid method of Refs.~\cite{Andersen2000,Pavarini2004}, in which  Wannier functions are constructed from outward solutions of the radial Schr\"odinger equation and their energy derivatives on a chosen grid of energies. 

Another approach \cite{Grechnev2007} makes use of the fact that some DFT band structure techniques expand Kohn-Sham states $\psi_{\vk\nu}$ in a basis containing, among others, suitable "atomic-like" functions for a given correlated shell; such functions are then employed as a correlated-subspace basis. The somewhat older method of Ref.~\cite{Savrasov2004} writes the whole Hamiltonian (\ref{eq:H_DFTU}) using  atomic-like basis functions instead of $\psi_{\vk\nu}$ and employs a subset of them to represent correlated orbitals; 
this approach is applicable only for few band-structure techniques employing such suitable basis functions.

Once the basis 	of correlated orbitals $w_{i\alpha}$ is chosen one needs to determine the on-site Coulomb repulsion between them. In principle, one may easily evaluate matrix elements of the bare Coulomb interactions $u(\vr)=1/r$ 
between such orbitals. The bare Coulomb repulsion is, however, known to be a very poor approximation for the local interaction in solids in eq.~\ref{eq:H_DFTU}.
For example, the average over its matrix elements between Ni 3$d$ orbitals in NiO evaluates to about 20$-$25~eV~\cite{Sakuma2013}. Experimentally, though, one finds that the splitting between  occupied and empty 3$d$ localized features seen in the PES/inverse-PES spectra, which is, to a first approximation, the average $\langle U\rangle$, amounts only to about 9~eV~\cite{Reinert2005}\footnote{The optical gap of about 4~eV in this compounds is of the charge-transfer (O~2$p \to$ Ni~3$d$) type.}. 
This discrepancy is, of course, due to the fact that the on-site interaction between localized orbitals in solids is strongly screened by itinerant states. Hence, one should view the Hamiltonian (\ref{eq:H_DFTU}) as a {\it low-energy} description of the correlated system, where  the interactions between localized states $w_{i\alpha}$ and itinerant bands, which are not explicitly included, have been integrated out. In result, the effective Coulomb repulsion $u(\vr,\vr',\omega)$ acquires  a frequency dependence, which is then passed to matrix elements in the correlated-orbitals basis:
\begin{equation}\label{eq:U_matel}
\langle 1 2|U| 3 4\rangle  (\omega)=\int d\vr d\vr' w_{i1}^*(\vr) w_{i2}^*(\vr')u(\vr,\vr',\omega)  w_{i3}(\vr) w_{i4}(\vr'),
\end{equation}
with the low-frequency limit of $\langle 1 2|U| 3 4\rangle  (\omega)$ giving a value of on-site repulsion that is strongly reduced by screening; it is relevant for the low-energy physics described by (\ref{eq:H_DFTU}). The high-frequency limit of $\langle 1 2|U| 3 4\rangle  (\omega)$ approaches the bare Coulomb value; this high-frequency tail of  $\langle 1 2|U| 3 4\rangle  (\omega)$ may affect the low-energy physics producing an additional enhancement of  quasiparticle renormalization~\cite{Casula2012}; it also induces  high-energy plasmonic spectral features~\cite{Casula2012b}. 

Due to this complex effect of screening the local Coulomb repulsion is rather difficult to evaluate from first principles and often treated as a parameter.
A  more consistent and truly {\it ab initio} approach is based on evaluating the screening  of local repulsion between a given set of local orbitals $w_{i\alpha}$ from the Kohn-Sham band structure. One popular approach of this kind, the constrained random-phase approximation (cRPA) \cite{Aryasetiawan2004}, separates the polarization function $\Pi(\omega)=\Pi_c(\omega)+\Pi_r(\omega)$ evaluated within RPA into the contribution $\Pi_c(\omega)$ due to transitions  within the subset of  correlated bands $\cal{W}$ and $\Pi_r(\omega)$ due to all other transitions.  Then the relevant  interaction  is obtained by screening the bare Coulomb repulsion $v(\vr)$ with $\Pi_r$
and then projecting $u(\vr,\vr',\omega)$ into the subspace of $w_{i\alpha}$ using (\ref{eq:U_matel}). The cRPA method is a powerful technique that is able to obtain all matrix elements of $\langle 1 2|U| 3 4\rangle  (\omega)$ with their frequency dependence. 
However, cRPA is not particularly well suited for the case of a significant entanglement between the correlated $\cal{W}$ and itinerant band subspaces, which is precisely the case in 3$d$ transition metals, where the dispersive 4$s$ band crosses and mixes with the narrow 3$d$ one. It is difficult to define a consistent separation of the polarization into $\Pi_c(\omega)$ and $\Pi_r(\omega)$ in this case, though some versions of cRPA to handle this entanglement have been formulated~\cite{Miyake2009,Seth2017}.

An alternative approach to first-principles evaluation of the local interaction is based on the assumption that a quantitatively correct static screening of the on-site interaction is already included at the DFT level through the local XC potential. This approach named constrained LDA (cLDA) \cite{Dederichs1984,Hybertsen1989,Cococcioni2005} constrains the charge on the localized shell of interest on a single site within a supercell with other states unconstrained, hence, allowed to screen the on-site interaction. The band energy of  corresponding "constrained" KS states is then evaluated as a function of the shell occupancy and its spin polarization in order to extract the direct Coulomb repulsion parameter $U$ and Hund's rule coupling $J_H$, respectively. The method was shown to provide reasonable values of the static interaction for pure iron \cite{Cococcioni2005, Belozerov2014}, though it is not free from uncertainties. 

The Kohn-Sham band structure, which is the quadratic part of the DFT+U Hamiltonian (\ref{eq:H_DFTU}), is that of non-interacting electrons moving in an effective potential. However, this potential contains, among other terms,  the Hartree and XC potentials corresponding to the electron density of the Kohn-Sham states. Hence, the Kohn-Sham bands are not truly those of a non-interacting system. In particular, the effect of the screened Coulomb interaction $u(\vr,\vr',\omega)$ acting between correlated orbitals is included
in a static mean-field way by standard DFT; this fact is used by the cLDA method described above to extract  the value of this interaction. As the same interaction explicitly enters  into (\ref{eq:H_DFTU}), it is necessary to remove this static mean-field contribution from the same Hamiltonain to avoid counting it twice. Hence, the corresponding double-counting correction (DC) is included as the last term into (\ref{eq:H_DFTU}).

Though the local screened interaction is certainly included  in some form by XC potentials determining its exact contribution is  a highly nontrivial  problem. Local and semi-local XC potentials are functions of the full charge density 
and also non-linear; they cannot be represented as a superposition of contribution due to different orbitals. Hence, the problem of formulating a theoretically sound expression for the DC  term has not been fully solved to date. There exist a number of different DC formulae~\cite{Anisimov1991,Czyzyk1994,Lichtenstein2001,Park2014,Haule2015b}. The most widely used ones are derived by assuming that XC potentials include  the local Coulomb interaction in an orbitally-independent form. That form is given by the  Hartree-Fock potential due to the on-site interaction term in (\ref{eq:H_DFTU}) for a particular limit of the correlated-shell occupancy matrix. It is assumed to be uniform within the "around-mean-field" (AMF) approach \cite{Anisimov1991}, which is usually employed for weakly and moderately-correlated metals, e.~g., in the case of iron. The alternative "fully-localized-limit' (FLL) form \cite{Czyzyk1994} assumes the most non-uniform occupancy matrix for a given shell filling and is generally employed for strongly-correlated systems like Mott insulators. The contribution due to this term into the one-electron potential for a given orbital $\alpha$ is given by
\begin{equation}\label{eq:Sigma_DC}
\Sigma_{DC}^{\alpha}=\frac{\partial E_{DC}}{\partial \rho_{\alpha}}|_{\hat{\rho}_{DC}},
\end{equation} 
where the derivative over the orbital occupancy $\rho_{\alpha}$ is taken at the shell's occupancy matrix $\hat{\rho}_{DC}$ corresponding to a given limit (AMF, FLL, etc.).


\subsection{Dynamical mean-field theory}\label{sec:dmft}

Once all terms in the Hamiltonian (\ref{eq:H_DFTU}) are determined the next step is, obviously, solving it to obtain the ground-state and excited properties of a given real system. This represents a formidable problem, as one may notice that this Hamiltonian  can be viewed as a multi-band generalization of the  famous one-band Hubbard model (HM)
for which no exact solution is known for the relevant 2d and 3d cases. A breakthrough in the study of HM was achieved in the beginning of 90th in the framework of dynamical mean-field theory (DMFT) \cite{Metzner1989,Georges1992,Georges1996}. Though initially the DMFT formalism was written for the one-band HM, here we present its formulation for the Hamiltonian (\ref{eq:H_DFTU}) in view of applications to realistic materials.  The DMFT framework focuses on the one-electron Green's function (GF)  defined  in the Kohn-Sham space and imaginary-time domain\footnote{The imaginary  time/frequency domain is often used in DMFT calculations for the technical reasons outlined in Sec.~\ref{sec:QIP}, though it is not necessary.}
as $G_{\nu\nu'}(\vk,\tau-\tau')=-\langle \text{T}[c_{\vk\nu}(\tau)c_{\vk\nu'}^{\dagger}(\tau')]\rangle$, where  $\text{T}$ is the time-ordering operator. Its Fourier transform $G(\vk,i\omega_n)$ is the GF in the imaginary-frequency domain, where $i\omega_n=i\pi(2n-1)T$ is the fermionic Matsubara grid for the temperature $T$. Correlation effects arising due to the interaction $U$ term of (\ref{eq:H_DFTU}) are encoded in the Kohn-Sham space by the electronic self-energy $\Sigma^{KS}(\vk,i\omega_n)=\hat{P}^{\dagger}(\vk)\Sigma(\vk,i\omega_n)\hat{P}(\vk)$ , where $\hat{P}(\vk)$ are projector matrices (\ref{eq:WF})  to the correlated subspace, $\Sigma(\vk,i\omega_n)$ is the self energy in that subspace spanned by the localized orbitals (\ref{eq:WF}). The interacting lattice GF is thus obtained by inserting $\Sigma(\vk,i\omega_n)$ through the Dyson equation:
\begin{equation}\label{eq:dyson}
G^{-1}(\vk,i\omega_n)=G_0^{-1}(\vk,i\omega_n)-\hat{P}^{\dagger}(\vk)\left(\Sigma(\vk,i\omega_n)-\Sigma_{DC}\right)\hat{P}(\vk), 
\end{equation}
into the non-interacting lattice GF $G_0$ given by the first term of (\ref{eq:H_DFTU}), with the DC for the self-energy defined by (\ref{eq:Sigma_DC}).

\begin{figure}[!htbp]
	\begin{center}
		\includegraphics[width=0.85\textwidth]{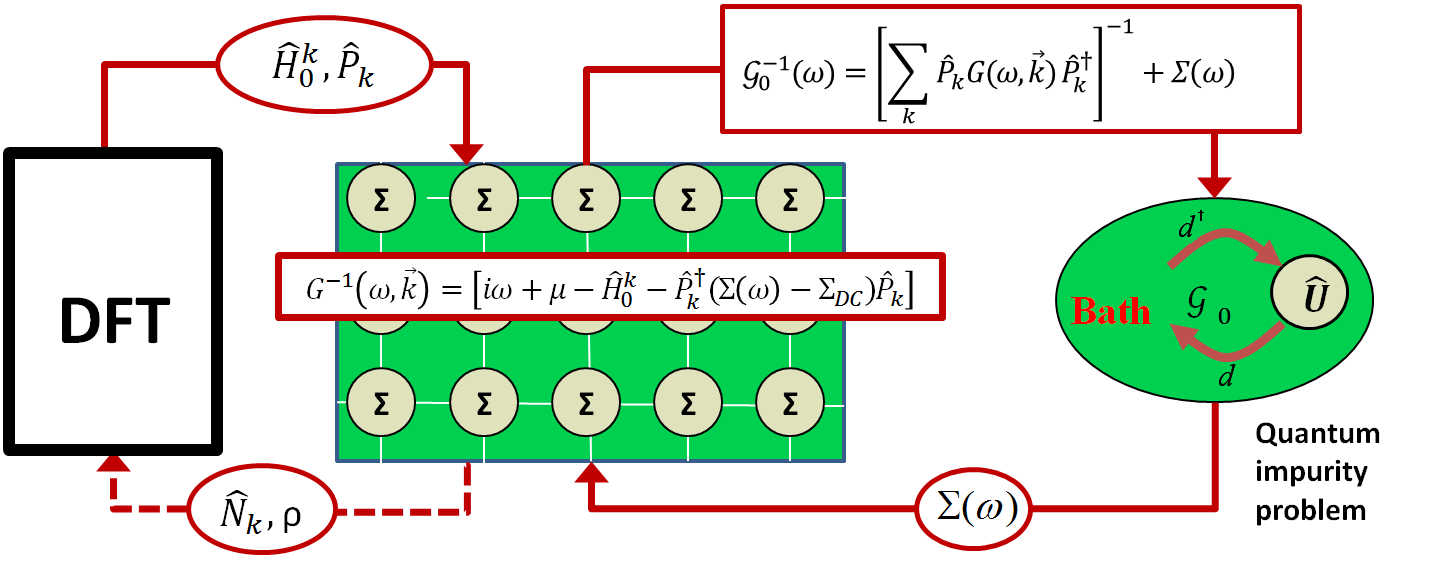}
	\end{center}
	\caption{Schematic diagram of the DFT+DMFT method. The initial input from the DFT part is the quadratic KS Hamiltonain $\hat{H}_0$ and projectors $\hat{P}$ between  the KS space and correlated subspace.  The right-hand side represents the DMFT cycle with the lattice problem mapped into the quantum-impurity one using eq.~\ref{eq:bathGF}; the calculated  impurity self-energy subsequently  is inserted back to the lattice, eq.~\ref{eq:dyson}. The updated DMFT density matrix can be inserted back to the DFT part (dashed arrow) to take into account modifications of the charge density and, therefore, $\hat{H}_0$, due to  correlations; this results in a DFT+DMFT framework that is self-consistent in the charge density.}
	\label{fig:DFT_DMFT_chart}
\end{figure}

The DMFT is based on the key observation of  Ref.~\cite{Metzner1989} that one may define a (non-trivial) infinite-dimensional  limit of (\ref{eq:H_DFTU}), and that the electronic self-energy becomes purely local in this limit, i.~e., $\vk$-independent\footnote{In the case of DFT+U Hamiltonian (\ref{eq:H_DFTU}) this approximation is applied to the self-energy $\Sigma(\vk,i\omega_n)$ in  the correlated subspace, while $\Sigma^{KS}$  can still be $\vk$ dependent due to the projectors $\hat{P}(\vk)$.},  $\Sigma(\vk,i\omega_n) \xrightarrow{d \rightarrow \infty} \Sigma(i\omega_n)$. Such single-site self-energy is given by the summation over irreducible (skeleton) Feynman diagrams involving only the single-site GF and the local vertex $\hat{U}$.  The coupling between a representative correlated shell $o$ and an effective electronic "bath" representing the rest of system is then given by the bath Green's function:
\begin{equation}\label{eq:bathGF}
\mathcal{G}_0^{-1}(i\omega_n)=\left[\sum_{\vk} \hat{P}(\vk)G(\vk,i\omega_n)\hat{P}^{\dagger}(\vk)\right]^{-1}+\Sigma(i\omega_n)=i\omega_n-\hat{\epsilon}-\Delta(i\omega_n),
\end{equation}
where $\hat{\epsilon}$ are bare (non-interacting)  single-site level positions, 
$\Delta(i\omega_n)$ is the hybridization function due to hopping between the site and electronic bath. The single-site problem in the correlated subspace is completely defined by (\ref{eq:bathGF}) and on-site Coulomb replusion 
$$
\hat{H}_U^{(o)}=\sum_{\atop 1,2,3,4}\langle 1 2|U| 3 4\rangle d^{\dagger}_{1} d^{\dagger}_{2} d_{4} d_{3}
$$ 
(omitting the irrelevant site index $o$ in $d$ and $d^{\dagger}$). The lattice problem is thus mapped into an auxilary {\it quantum impurity problem} (QIP) \cite{Georges1992} for a single correlated shell, which is fully analogous to the standard Anderson impurity model (AIM). However, in contrast to the usual AIM, $\Delta(i\omega_n)$ is not given by the hybridization of non-interacting bands; it should be rather viewed as a dynamical mean-field  implicitly depended on the single-site self-energy through eqs. (\ref{eq:dyson}-\ref{eq:bathGF}). By solving the QIP, i. e., by summing (all or subset of) local Feynman diagrams one obtains the impurity GF and self-energy:
\begin{equation}\label{eq:QIP}
\left\{\mathcal{G}(i\omega_n), \hat{H}_U^{(o)}\right\}  \rightarrow \left\{G_{imp}(i\omega_n),\Sigma_{imp}(i\omega_n) \right\}.
\end{equation}
One then employs the standard recipe to close   the mean-field cycle as shown in Fig.~\ref{fig:DFT_DMFT_chart}: the obtained impurity self-energy is inserted for all correlated shells to restore the translational invariance, $\Sigma(\vk,i\omega_n)\equiv \Sigma_{imp}(i\omega_n)$ allowing to update the chemical potential $\mu$ and to recalculate the mean field $\mathcal{G}_0$ by eqs. (\ref{eq:dyson}-\ref{eq:bathGF}). This cycle is iterated until the self-consistency is reached: the QIP solved for the mean-field $\mathcal{G}_0$ results in the same self-energy $\Sigma$ that was used to obtain this mean-field through (\ref{eq:dyson}-\ref{eq:bathGF}). Alternatively, the same self-consistency condition is represented by $G_{imp}(i\omega_n)=G_{loc}(i\omega_n)$, where
\begin{equation}\label{eq:G_loc}
G_{loc}(i\omega_n)=\sum_{\vk}\hat{P}(\vk)G(\vk,i\omega_n)\hat{P}^{\dagger}(\vk)
\end{equation}
is the local GF of lattice problem. The problem defined by the Hamiltonian (\ref{eq:H_DFTU}) is thus exactly solved in the limit of infinite lattice connectivity, as can be also shown explicitly, see \cite{Georges1996}. As for any mean-field approach the usefulness of DMFT method is based on its ability to describe  the realistic  3d lattices, for which the single-site approximation  $\Sigma(\vk,i\omega_n) \rightarrow \Sigma(i\omega_n)$ appears to be rather reasonable, though it is not quantitatively exact. At the same time the single-site dynamics due to electronic correlations is fully included in DMFT;  this explains its success in reproducing such non-perturbative phenomena as the Mott transition. The method captures not only the insulating $U/W \to \infty$ and non-interacting $U/W \to 0$ limits (where $W$ is the bandwidth of non-interacting bands $\epsilon_{vk}$ in (\ref{eq:H_DFTU}))  but also all intermediate regimes given by finite $U/W$.

For 2d and quasi-2d systems the single-site DMFT is generally not an adequate approximation. The $\vk$ dependence of the self-energy is key to describe, for example, the physics of layered cuprate superconductors, in particular, their PES \cite{Damascelli2003}. This problem was addressed by cluster extensions of the single-site DMFT, which were formulated in both the real and reciprocal spaces~\cite{Potthoff2003,Maier2005,Ferrero2009}. The single-site QIP  (\ref{eq:QIP}) is thus generalized to the corresponding cluster problem. Such generalization increases dramatically the computational cost of solving the QIP, hence, the cluster methods are not generally applicable to full $d$ and $f$ shells; they have been  extensively applied to quasi-1band systems like layered cuprates. Another more recent effort in development of extended-DMFT frameworks \cite{Rubtsov2008,Held2008,Ayral2015} is based on applying the single-site approximation to two-electron correlation functions (like the vertex function) while keeping the $\vk$-dependence of the one-electron self-energy. These approaches are promising for applications to multiband systems, though they are still currently too heavy for applications in the cases considered in this highlight, when many-electron effects for the full $d$ shell need to be taken into account. 

\subsection{The quantum impurity problem}\label{sec:QIP}

The QIP problem schematically given by eq.~\ref{eq:QIP} is a true many-electron problem, though a single-site one, and represents, in fact, a numerical "bottleneck" of the DFT+DMFT framework. In the imaginary-time path integral formalism (see e.g. \cite{Negele1988}) it reads
\begin{equation}\label{eq:int_G}
G_{\alpha\alpha'}(\tau_0-\tau_1)=\frac{1}{Z}\int \mathcal{D}[d,d^{\dagger}]d_{\alpha}(\tau_0)d_{\alpha'}^{\dagger}(\tau_1)\exp[-S],
\end{equation}
where $\mathcal{D}[d,d^{\dagger}]$ is the path integration over all impurity degrees of freedom and 
\begin{equation}\label{eq:int_Zpath}
Z=\int \mathcal{D}[d,d^{\dagger}]\exp[-S]
\end{equation}
is the impurity partition function, $S$ is the impurity action:
\begin{equation}\label{eq:S_int}
S=\sum_{\alpha_1\alpha_2}\int d\tau \int d\tau' d_{\alpha_1}^{\dagger}(\tau)\left[\mathcal{G}_0^{-1}(\tau-\tau')\right]_{\alpha_1\alpha_2}d_{\alpha_2}(\tau')+\int d\tau \hat{H}_U^{(o)}(\tau).
\end{equation}
Many-body methods to evaluate (\ref{eq:int_G}) represent a large research field   initiated by early studies of AIM and very actively developed at present, in particular, to provide efficient quantum-impurity "solvers" for the DMFT framework. They will not be reviewed here in any details; we will only briefly outline main strategies for solving the QIP and provide some useful references. 

The methods dealing with QIP can be divided into numerically-exact and approximate analytical kinds. Among the former one may especially mention stochastic quantum Monte Carlo (QMC) methods; a breakthrough in this domain   has been achieved by so-called "continuous-time" (CT) QMC methods \cite{Prokof'ev1998} (see review~\cite{Gull2011} on its applications to the fermionic QIP). The most popular  CT-QMC approaches are based on an expansion of the partition function (\ref{eq:int_Zpath}) in powers of $\hat{H}_U^{(o)}$~\cite{Rubtsov2005} or, alternatively, in powers of the hybridization function $\Delta(\tau)$ \cite{Werner2006}, see eq.~\ref{eq:bathGF}. One subsequently sums up various diagrammatic contributions into GF (\ref{eq:int_G}) and other correlation function  in accordance with their relative weight in $Z$ by employing a Monte Carlo importance sampling. In contrast to the older QMC approach of  Hirch and Fye \cite{Hirsch1986} based on discretization of the integrals over $\tau$ in (\ref{eq:S_int})  the CT-QMC approach is free from the discretization error and can treat more complex interaction vertices  $\hat{H}_U^{(o)}$. All these QMC methods generally work in the imaginary-time/imaginary-frequency domain, hence, the resulting GF needs to be analytically continued to the real-energy axis to obtain an experimentally-observable real-frequency spectra.

The hybridization-expansion CT-QMC technique employed as a quantum-impurity solver in the DFT+DMFT calculations presented in this review. This approach is sufficiently computationally efficient to solve the QIP for the whole Fe 3$d$ shell. Particularly, the case of simplified, "density-density" Coulomb vertex $\hat{H}_U^{(o)}$ reducible to the form  $\sum_{\alpha \alpha'}U_{\alpha\alpha'}\hat{n}_{\alpha}\hat{n}_{\alpha'}$   allows to employ the fast "segment-picture" algorithm \cite{Werner2006,Gull2011}, reducing the computational effort very significantly. The density-density approximation neglects some potentially important matrix elements of the Coulomb vertex\footnote{For example, the "spin-flip" contributions  to $\hat{H}_U^{(o)}$ of the form $d^{\dagger}_{m\uparrow} d^{\dagger}_{m'\downarrow}d_{m'\uparrow} d_{m\downarrow}$ cannot be reduced to a density-density form.} and thus introduces a system-dependent error.  In the case of moderately-correlated metal like iron it does not affect  the qualitative picture, but is still quantitatively important (see Appendix~\ref{chap:appendix_FeDensDens}); for strongly-correlated systems as, for example, FeSe \cite{Aichhorn2010} this approximation may lead to qualitatively wrong results. Calculation with the full 4-index vertex are much more computationally demanding, but still nowdays possible thanks to a recent development of fast algorithms \cite{Lauchli2009,Gull_thesis}. 

Another popular numerically-exact approach, the exact diagonalization technique \cite{Caffarel1994}, see also \cite{Lu2014,Go2017} for more recent developments. It is based on representing the hybridization function by a set of auxiliary discrete levels $\{\epsilon_b\}$ of the bath mixing with the impurity states, $\Delta_{\alpha\alpha'}(\omega) \sim \sum_{b}\frac{V_{b\alpha}V_{b\alpha'}^{\dagger}}{\omega-\epsilon_b}$. The resulting large Hamiltonian including both impurity and bath states is subsequently diagonalized by Lanczos or similar techniques allowing to compute the impurity GF from obtained eigenvalues and eigenstates. 
Among the exact methods one should also mention the numerical renormalization-group and density-matrix renormalization-group methods \cite{Bulla2008,Hallberg2006}.

Analytical approaches are generally applicable only in certain regimes (strong or weak coupling). Weak-coupling methods are suitable for metallic phases; they are based on the standard Wick theorem and subsequent summation of a certain subset of Feynmann diagrams, like, for example, the fluctuation-exchange approximation \cite{Katsnelson1999,Drchal2005,Pourovskii2005}, which has been extensively applied to spectral properties of iron and nickel \cite{Braun2006,Grechnev2007,Sanchez-Barriga2009,Minar2014}.
Among numerous other analytical methods one may also mention the "slave" particle  approach \cite{Kotliar1986,deMedici2005,Lechermann2007,Rohringer2018} providing an economical and numerically efficient treatment  of the quasiparticle renormalization in multiband systems. The obvious advantage of these analytical techniques is their computational efficiency. They can also easily evaluate the GF and, hence, the measurable one-electron spectra, at the real-frequency axis.

Finally, the simplest approach to solving the QIP consists in employing the static Hartree-Fock approximation; in this case DFT+DMFT  is reduced to the popular LDA+U method \cite{Anisimov1991,Anisimov1997}.

\subsection{Charge density and total energy}\label{sec:etot}

As a result of the DMFT cycle (Fig.~\ref{fig:DFT_DMFT_chart}) one obtains the converged interacting lattice GF (\ref{eq:dyson}) in the KS space. The corresponding density matrix 
\begin{equation}\label{eq:Nk}
N^{\vk}_{\nu\nu'}=\sum_n G_{\nu\nu'}(\vk,i\omega_n)e^{i\omega_n0^+}
\end{equation}
gives the contribution of KS bands in $\mathcal{W}$ to the charge density. Therefore, the charge density $n(\vr)$ is affected by many-electron effects through the DMFT self-energy $\Sigma(i\omega_n)$ entering into $G(\vk,i\omega_n)$; the KS one-electron potential being a functional of $n(\vr)$ is thus modified as well. Hence, the one-electron part $H_0$ of the DFT+U Hamiltonain (\ref{eq:H_DFTU}) comes out to be implicitly dependent on $\Sigma(i\omega_n)$. 

This observation led to formulation of the charge self-consistent DFT+DMFT framework, in which  $n(\vr)$ and $H_0$ are consistently updated  to take into account the impact of correlations as shown in the left-hand side of Fig.~\ref{fig:DFT_DMFT_chart}. In practice, $\hat{N}^{\vk}$ in the KS basis is submitted back to the DFT part; the corresponding contribution to $n(\vr)$ is then calculated  through the expansion of $\psi_{\vk\nu}$ in the basis of a given band-structure method.  Several such self-consistent DFT+DMFT frameworks have been implemented recently \cite{Savrasov2004,Minar2005,Pourovskii2007,Haule2010,Aichhorn2011,Granas2012,Park2014_1}.

In this self-consistent framework the DMFT self-consistency condition, $G_{loc} \equiv G_{imp}$, as well as the relation between the  KS potential and electronic density are derived by extremization of the following DFT+DMFT grand potential \cite{Kotliar2006} :

\begin{align}\label{dmft_gp}
\Omega\left[n({\bf r}),G_{loc},\Delta \Sigma,\hat{\epsilon}\right]=&-\Tr \ln\left[i\omega_n+\mu-H_0- \Delta\Sigma
\right]  -\Tr\left[G_{loc}\Delta \Sigma\right] \\
& + \sum_{{\bf R}}\left[\Phi_{imp}[G_{loc}({\bf R})]-\Phi_{DC}[G_{loc}({\bf R})]\right]+\Omega_{r}[n({\bf r})] \nonumber \\
& \equiv\Delta \Omega\left[G_{loc},\Delta \Sigma,V_{KS}\right]+\Omega_{r}[n({\bf r})] \nonumber ,
\end{align}
where $\Delta\Sigma$ is the difference between the impurity self-energy $\Sigma_{imp}$ and the double counting correction (\ref{eq:Sigma_DC}), $\Phi_{imp}[G_{loc}({\bf R})]$ is the  DMFT interaction energy functional for the site 
${\bf R}$,   $\Phi_{DC}[G^{loc}_{\bf R}]$ is the corresponding functional for the double-counting correction. The last term $\Omega_{r}[n({\bf r})]$ depends only on the electronic charge density $n({\bf r})$ and comprises the electron-nuclei, Hartree and exchange-correlation contribution, while all other terms collected in $\Delta \Omega\left[G_{loc},\Delta \Sigma,V_{KS}\right]$ do not have an explicit dependence on $n({\bf r})$.   
From the zero-temperature limit of (\ref{dmft_gp})  one derives \cite{Amadon2006} the following expression for the total energy: 
\begin{equation}\label{eq:E_DMFT}
E_{DFT+DMFT}=\sum_{\vk\nu} \epsilon_{\vk\nu} N_{\nu\nu}^{\vk} +\langle H_U \rangle -E_{DC} +E_{en}[n(\vr)]+E_H[n(\vr)]+E_{xc}[n(\vr)],
\end{equation}
where $E_{en}$, $E_H$, $E_{xc}$ are the standard DFT electron-nuclei, Hartree and exchange-correlation contributions evaluated from the charge density $n(\vr)$ that includes the DMFT correction. The interaction energy $\langle H_U \rangle$ can be evaluated from the self-energy using the Migdal formula $\langle H_U \rangle= \frac{1}{2}\Tr\left[\Sigma_{imp} G_{imp}\right]$, alternatively, the expectation value $\langle d^{\dagger}_{1} d^{\dagger}_{2} d_{3} d_{4}\rangle$ can be directly measured, e. g., by  using QMC quantum-impurity solvers.

Instead of the self-consistent charge density $n(\vr)$ one may employ in (\ref{eq:E_DMFT}) the DFT one, $n_{DFT}(\vr)$ resulting in the so-called "one-shot DMFT" scheme. The impact of the self-consistency in charge density on the DFT+DMFT total-energy and spectra has been studied in a number of works \cite{Pourovskii2007,Aichhorn2011,Amadon2012,Leonov2015,Bhandary2016}, though a consistent assessment for the full range of correlation strength is still lacking. However, the charge-density self-consistency seems to important for localized systems as $\gamma$-Ce and Ce oxides \cite{Pourovskii2007} and VO$_2$ \cite{Leonov2015}. The possible reason pointed out by Ref.~\cite{Bhandary2016} is that the occupancy of $\psi_{\vk\nu}$ states is very different in the localized limit as compared to a metallic band structure  predicted by DFT.  In the former case the KS states $\vk\nu$ of correlated bands will be all roughly half-filed due to the contribution of corresponding lower Hubbard band. In DFT  the KS states $\vk\nu$ are  occupied below $E_F$ and empty above, hence, the occupancy varies strongly in the $\vk$ space. Another important effect of the charge-density self-consistency is an overall lower sensitivity of the result to the choice of DC; changes in DC seem to be compensated by the corresponding  modifications in $V_{KS}$ \cite{Aichhorn2011}.

\newpage
\section{The impact of density-density approximation: a benchmark}
\label{chap:appendix_FeDensDens}

\begin{figure}[!b]
	\includegraphics[width=0.45\textwidth]{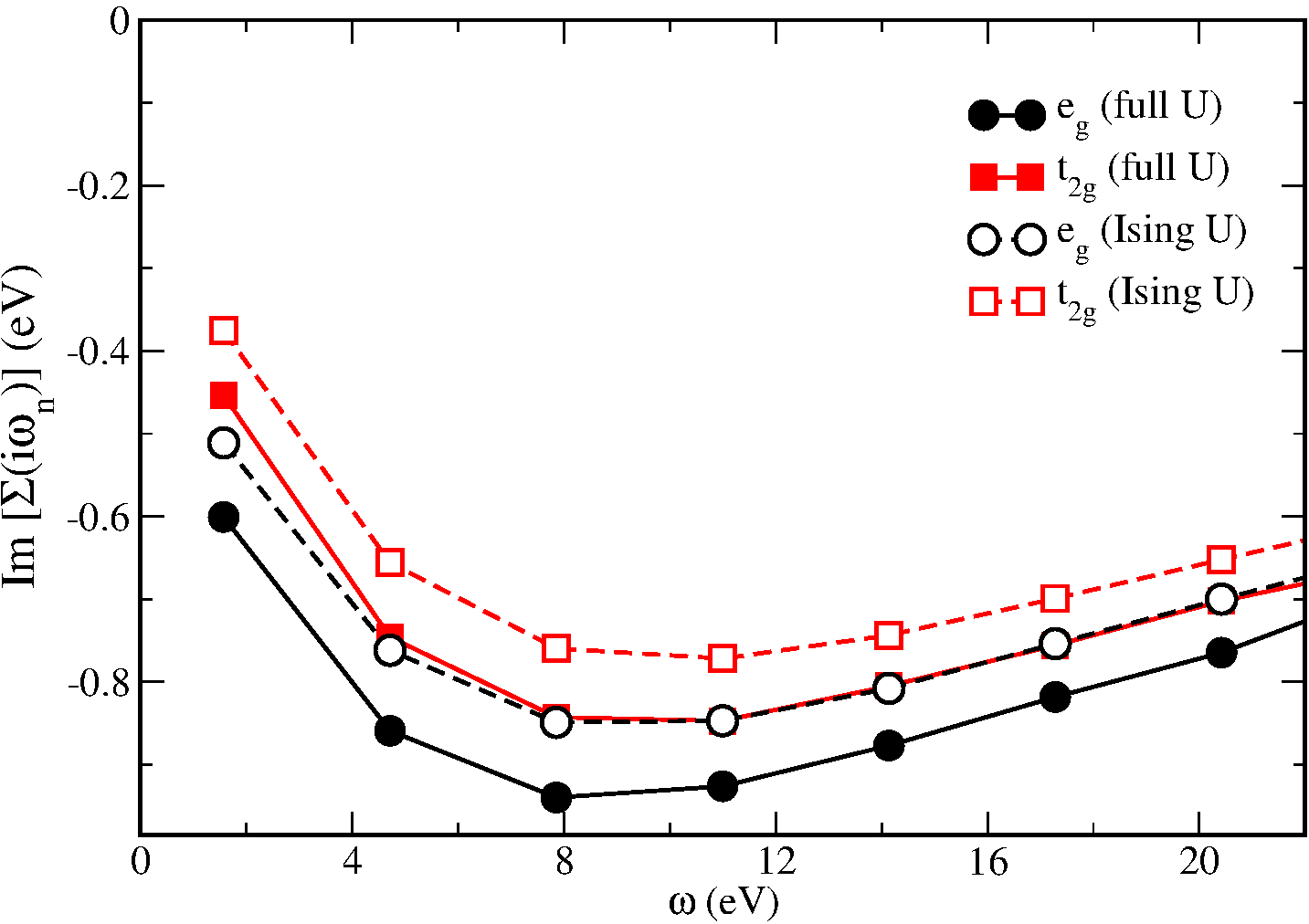}
	\includegraphics[width=0.45\textwidth]{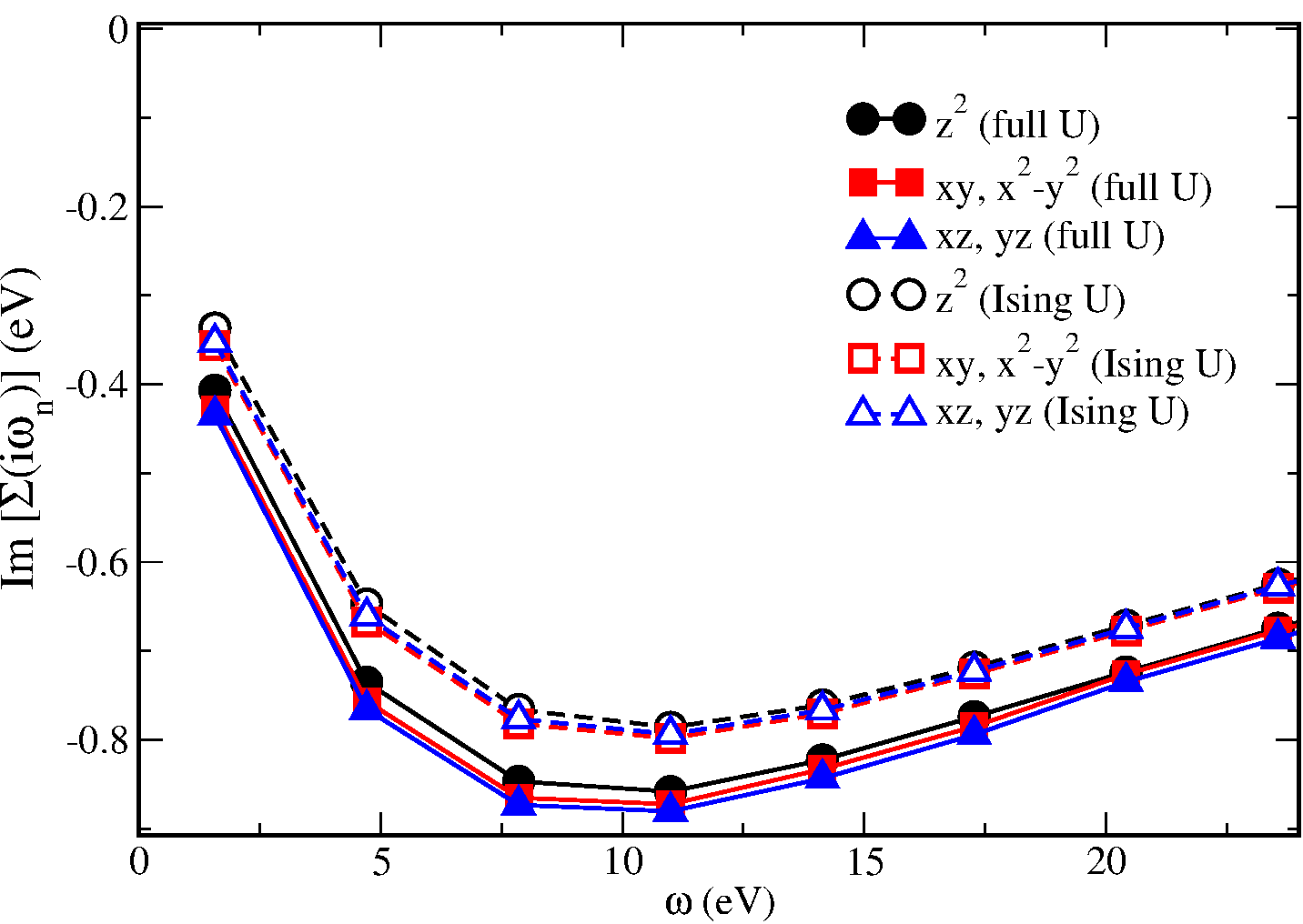}
	\caption{Left panel:  The imaginary part of DMFT self-energy on the Matusbara grid for the non-degenerate orbitals of the Fe 3$d$ shell in the bcc structure calculated with the rotationally-invariant (filled symbols) and  density-density (empty symbols) local Coulomb interaction, respectively. Right panel: the same for the hcp structure.}
	\label{fig:self-bcc_hcp}
\end{figure}

In this appendix we illustrate the impact of density-density approximation for the local Coulomb interaction by performing DFT+DMFT calculations with and without this approximation for the bcc $\alpha$ and hcp $\epsilon$ iron phases at the Earth's core condition. Self-consistent in the charge density DFT+DMFT calculations (Sec.~\ref{sec:etot}) were thus carried out for the perfect bcc and hcp lattices at the atomic volume of 7.05 \AA$^3$/atom expected for the inner core of Earth and the temperature of 5800~K. The on-site Coulomb interaction was defined by the parameters $U=$5.0~eV, $J_H=$0.93~eV previously used in the study of $\epsilon$-Fe by Ref.~\cite{Pourovskii2017}; the same choice for  the energy window ( [-10.8 eV, 4.0 eV] around the Fermi level) was also employed for the Kohn-Sham states used to construct Wannier orbitals representing Fe 3$d$ states.  The DMFT impurity problem was solved by the hybridization-expansion quantum Monte Carlo impurity solver using its segment-picture version \cite{Werner2006,Gull2011} in the case of density-density (Ising) vertex and the implementation of Seth {\it et al.}~\cite{Seth2016} in the case of full rotationally-invariant one.

The resulting DMFT self-energies for both phases are compared in Fig.~\ref{fig:self-bcc_hcp}. For both bcc and hcp-Fe the use of density-density approximation results in a systematic underestimation of the magnitude of scattering  $|Im \Sigma(i\omega_n)|$, which is, however, more pronounced in the case of more correlated bcc.  Qualitative features, like the $e_g$ orbitals  markedly more correlated than the $t_{2g}$ ones in bcc-Fe as well as a uniform Fermi-liquid behavior of all orbitals in hcp, are well captured within the density-density approximation. We have also  calculated the transport using the approach outlined in Sec.~\ref{subsec:epsFe_transport} and these self-energies analytically continued to the real-energy axis.  The electrical and thermal conductivities for bcc Fe are found to be overestimated by 40\% and 29\%, respectively, due to the density-density approximation. As expected, the impact of this approximation for the less-correlated $\epsilon$-phase is smaller  and amounts to 33\% and 23\%, respectively. Hence, though the use of full vertex does not lead to qualitative changes it is still found to be important for quantitative results.

\bibliographystyle{iopart-num}
\providecommand{\newblock}{}

\end{document}